\documentclass{aa}
\usepackage{rotating}
\usepackage{graphicx}
\usepackage{txfonts}
\usepackage{natbib}
\newcommand{\stars}{HD~54662}
\newcommand{\rsun}{R_{\odot}}
\newcommand{\msun}{M_{\odot}}

\usepackage{pdfpages}

\begin{document}

\title{The long-period massive binary HD~54662 revisited\thanks{Based on observations collected with the TIGRE telescope (La Luz, Mexico) and at the European Southern Observatory (La Silla and Cerro Paranal, Chile), the Nordic Optical Telescope (La Palma, Spain) and the Canada-France-Hawaii Telescope (Mauna Kea, Hawaii).
Also based on observations collected with {\it XMM-Newton}, an ESA Science Mission with instruments and contributions directly funded by ESA Member States and the USA (NASA). 
}}
\author{E.\ Mossoux\inst{1} \and L.\ Mahy\inst{2,1}\fnmsep\thanks{FRS-FNRS Postdoctoral Researcher} \and G.\ Rauw\inst{1}}

\offprints{E.\ Mossoux}
\mail{@uliege.be}
\institute{Space sciences, Technologies and Astrophysics Research (STAR) Institute, Universit\'e de Li\`ege, All\'ee du 6 Ao\^ut, 19c, B\^at B5c, 4000 Li\`ege, Belgium
\and Instituut voor Sterrenkunde, KU Leuven, Celestijnenlaan 200D, Bus 2401, 3001 Leuven, Belgium}
\date{Received date / Accepted date}
\abstract{HD~54662 is an O-type binary star belonging to the CMa OB1 association. Because of its long-period orbit, this system is an interesting target to test the adiabatic wind shock model.}
{The goal of this study is to improve our knowledge of the orbital and stellar parameters of HD~54662 and to analyze its X-ray emission to test the theoretical scaling of X-ray emission with orbital separation for adiabatic wind shocks.}
{We applied a spectral disentangling code to a set of optical spectra to determine the radial velocities and the individual spectra of the primary and secondary stars. 
The orbital solution of the system was established and the reconstructed individual spectra were analyzed by means of the CMFGEN model atmosphere code. 
We fitted two X-ray spectra using a Markov Chain Monte Carlo algorithm and compared these spectra to the emission expected from adiabatic shocks.}
{We determine an orbital period of 2103.4~days, a surprisingly low orbital eccentricity of 0.11, and a mass ratio $m_2/m_1$ of 0.84.
Combined with the orbital inclination inferred in a previous astrometric study, we obtain surprisingly low masses of 9.7 and $8.2\,\msun{}$.
From the disentangled primary and secondary spectra, we infer O6.5 spectral types for both stars, of which the primary is about two times brighter than the secondary. 
The softness of the X-ray spectra for the two observations, the very small variation of best-fitting spectral parameters, and the comparison of the X-ray-to-bolometric luminosity ratio with the canonical value for O-type stars allow us to conclude that X-ray emission from the  wind interaction region is quite low and that the observed emission is rather dominated by the intrinsic emission from the stars. We cannot confirm the runaway status previously attributed to HD~54662 by computing the peculiar radial and tangential velocities.
We find no X-ray emission associated with the bow shock detected in the infrared.}
{The lack of hard X-ray emission from the wind-shock region suggests that the mass-loss rates are lower than expected and/or that the pre-shock wind velocities are much lower than the terminal wind velocities.
The bow shock associated with HD 54662 possibly corresponds to a wind-blown arc created by the interaction of the stellar winds with the ionized gas of the CMa OB1 association rather than by a large differential velocity between the binary and the surrounding interstellar medium.}

\authorrunning{E. Mossoux et al.}
\titlerunning{HD54662}
\maketitle
\section{Introduction}
Over recent years, the question of the incidence of binarity among Population\,I O-type stars has moved to the forefront of massive stars research. 
\citet{SF} claimed that more than 70\% of O stars are currently members of a binary (or higher multiplicity) system or have been members in the past. Yet, such a high percentage of multiples includes all kinds of systems, from very hard systems with short orbital periods of a few days (best studied via photometric eclipses and spectroscopy) up to weakly bound systems with orbital periods of decades or even centuries (best studied with interferometry). In between these two extremes lies a region of the parameter space that is currently the least known. In fact, very few O-type binaries are known to have orbital periods longer than a year, but shorter than a century. To a very large extent, this situation reflects an observational bias: none of the most commonly used techniques are very sensitive in this regime and only long-term spectroscopic monitoring has been able so far to partly fill the gap \citep[see, e.g.,\ the cases of Cyg\,OB2 \#9 and 9~Sgr,][]{YN12,9Sgr}.

An interesting target in this context is \stars{} \citep[O7\,Vz var?][]{Sota}, the brightest and earliest member of the CMa OB1 association \citep{Gies}. 
\stars{} is considered as a runaway star because it has apparently created a bow shock in the interstellar medium (ISM) \citep{Noriega,Peri}. 
The multiplicity of this star has been uncertain for many years until \citet{Fullerton} observed a partial line splitting, revealing the binary nature of the system. 
Based on {\it IUE} spectra, \citet{SL} suggested an orbital period of about 92 days. \citet{Boyajian} combined their own radial velocity (RV) measurements with data from the literature.
They derived a preliminary orbital solution with a period of 558\,days. 
However, as a result of the limited spectral resolution (R = 9000) of their data, they were only able to deblend the secondary signature  on mean spectra and noted that the semi-amplitude of their SB1 RV curve was very probably underestimated. 

\stars{} was resolved in interferometry with the VLTI and the PIONIER instrument \citep{SMASH}, and very recently, \citet{LeBouquin} proposed a full interferometric orbit of the system. 
They affirmed that the 558\,days orbital period proposed by \citet{Boyajian} is too short and rather favored a period of 2103\,days. A similar conclusion was reached by \citet{Sota} who quoted an unpublished study of Gamen et al.\ that indicates a period of 2119\,days. 

\stars{} is also an interesting target for the study of wind-wind interactions. 
Indeed, hot, massive stars blow powerful stellar winds, which generate huge mass-loss rates and large wind velocities. 
Whenever two massive stars are bound by gravity, their winds interact and this interaction produces signatures over a broad range of wavelengths. 
Among the most spectacular manifestations of this phenomenon is the X-ray emission from the shock-heated plasma within the wind interaction zone \citep[see][for a recent review]{AdSR}.
The plasma in the wind interaction region of long-period binaries is expected to be adiabatic. 
In such systems the winds of both stars have reached their terminal velocities before they collide. 
This leads to relatively low post-shock plasma densities and thus long cooling times. The X-ray emission of such an adiabatic interaction region should vary as the inverse of the distance $d$ between the stars \citep{stevens92}. Whilst this picture is consistent with the observations of Cyg\,OB2 \#9 \citep{YN12}, it fails in the case of 9~Sgr \citep{9Sgr}. 
However, the latter O + O binaries are also nonthermal radio emitters, which implies the presence of a population of relativistic electrons that could modify the properties of the wind interaction \citep{PittDough}, thereby causing deviations from the theoretical $1/d$ relation.
\stars{} should feature an adiabatic interaction region because of its long period, but has not been reported as a nonthermal radio emitter; this makes it an interesting target to test the $1/d$ relation.

\section{Observations}
\subsection{Optical spectroscopy}
Fifteen spectra were obtained between October 2014 and March 2017 with the 1.2~m TIGRE telescope \citep[][]{Hempelmann,Mittag,Schmitt} at La Luz Observatory near Guanajuato (Mexico). 
The TIGRE instrument is operated in a fully robotic way and features the refurbished HEROS \'echelle spectrograph \citep{Kaufer2}. 
The HEROS spectrograph offers a spectral resolving power of 20\,000 over the full optical range with a small gap near 5800\,\AA. The HEROS data were reduced with the corresponding reduction pipeline \citep{Mittag,Schmitt}. 

The HEROS data were complemented by \'echelle spectra extracted from several archives. Thirteen spectra were taken from the archive of the UVES spectrograph \citep{Dekker}. 
The UVES spectrograph is operated on the 8.2~m Very Large Telescope UT2 (ESO, Cerro Paranal) and offers a spectral resolving power near 40\,000 for a 1$\arcsec$ slit. 
Nine spectra were taken with the FEROS spectrograph \citep[$R \sim 48\,000$ over the full optical range;][]{Kaufer} on the 2.2~m ESO/MPI telescope at La Silla (ESO).
One spectrum was taken in the context of the IACOB project \citep{IACOB} with FIES at the 2.5~m Nordic Optical Telescope (Roque de los Muchachos, La Palma). 
The spectrograph was operated with fibre \#3 providing a resolving power of 46\,000. 
Finally, one spectrum was obtained from the Canada-France-Hawaii Telescope (CFHT) archive. 
The spectrum was taken with the ESPaDOnS spectropolarimeter \citep[$R \sim 68\,000$][]{Donati} at the 3.6~m CFHT (Mauna Kea, Hawaii).

For all optical spectra, we used the {\tt telluric} tool within IRAF\footnote{IRAF is distributed by the National Optical Astronomy Observatories, which are operated by the Association of Universities for Research in Astronomy, Inc., under contract with the National Science Foundation.} along with the list of telluric lines of \citet{Hinkle} to remove the telluric absorptions around the He\,{\sc i} $\lambda\lambda$\,5876, 7065, and H$\alpha$ lines. The spectra were continuum normalized using MIDAS routines and adopting the same set of continuum windows for all spectra to achieve self-consistent results. The full list of our 39 optical spectra is given in Table\,\ref{journalopt}.

\begin{table}
\caption{Journal of the optical spectroscopy observations of \stars{}.\label{journalopt}}
\begin{center}
\begin{tabular}{c c c l}
\hline
HJD-2450000 & RV$_1$ & RV$_2$ & Inst.\\
            & (km\,s$^{-1}$) &(km\,s$^{-1}$)& \\
\hline 
3701.8148  &   21.7  &  60.9  & UT2 + UVES \\
3739.6729  &   23.9  &  59.1  & MPI2.2 + FEROS\\ 
3740.7216  &   24.0  &  62.0  & MPI2.2 + FEROS\\ 
3740.7414  &   24.1  &  62.5  & MPI2.2 + FEROS\\
3773.7029  &   24.4  &  61.0  & MPI2.2 + FEROS\\
3773.7141  &   24.3  &  57.7  & MPI2.2 + FEROS\\
4600.4952  &   65.1  &  16.0  & MPI2.2 + FEROS\\
5145.7407  &   45.2  &  30.8  & NOT + FIES \\
5489.0901  &   24.9  &  50.1  & CFHT + ESPaDOnS \\
5606.5411  &   26.2  &  54.2  & MPI2.2 + FEROS\\
5664.5244  &   22.2  &  59.4  & UT2 + UVES \\
5667.4950  &   22.4  &  60.0  & UT2 + UVES \\
5696.4662  &   23.2  &  60.7  & MPI2.2 + FEROS\\
5799.8899  &   21.8  &  60.5  & UT2 + UVES \\
6008.5997  &   28.1  &  54.7  & UT2 + UVES \\
6067.4626  &   32.7  &  50.3  & MPI2.2 + FEROS\\
6239.8542  &   42.8  &  34.6  & UT2 + UVES \\
6259.7983  &   44.1  &  34.5  & UT2 + UVES \\
6275.7963  &   45.3  &  32.4  & UT2 + UVES \\
6305.8448  &   42.3  &  27.3  & UT2 + UVES \\
6317.5487  &   49.1  &  30.0  & UT2 + UVES \\
6336.5445  &   49.8  &  27.5  & UT2 + UVES \\ 
6946.9447  &   57.9  &  22.6  & TIGRE + HEROS \\
6975.9420  &   57.7  &  26.7  & TIGRE + HEROS \\
7006.8488  &   56.1  &  22.7  & TIGRE + HEROS \\
7033.8078  &   54.1  &  17.5  & TIGRE + HEROS \\
7034.7919  &   54.8  &  19.5  & TIGRE + HEROS \\
7058.6779  &   54.4  &  20.4  & TIGRE + HEROS \\
7074.6693  &   53.5  &  22.8  & TIGRE + HEROS \\
7085.7193  &   53.5  &  23.7  & TIGRE + HEROS \\
7100.6155  &   51.5  &  20.7  & TIGRE + HEROS \\
7333.9497  &   41.0  &  35.7  & TIGRE + HEROS \\
7340.8005  &   41.9  &  37.3  & UT2 + UVES \\
7347.7675  &   40.6  &  35.1  & UT2 + UVES \\   
7373.8459  &   39.1  &  39.7  & TIGRE + HEROS \\
7404.7724  &   36.9  &  39.6  & TIGRE + HEROS \\
7435.6917  &   35.1  &  42.0  & TIGRE + HEROS \\
7797.6935  &   22.9  &  52.9  & TIGRE + HEROS \\
7829.5827  &   24.2  &  57.8  & TIGRE + HEROS \\
\hline
\end{tabular}
\end{center}
\tablefoot{Radial velocities are determined with the spectral disentangling routine for the spectral regions from 4050 -- 4890\AA\ and around the He\,{\sc i} $\lambda$\,5876 line (see Sect.\,\ref{RVsol}). Typical uncertainties on RVs are 2\,km\,s$^{-1}$ for the primary and 5\,km\,s$^{-1}$ for the secondary.}
\end{table}

\subsection{{\it XMM-Newton} observations}
\label{xmm_obs}
\stars{} was observed with {\it XMM-Newton} \citep{jansen01} on 2014 October 1 and 2016 September 29. The journal of observations is given in Table~\ref{journal}.

For both observations, the EPIC cameras \citep{MOS,pn} were operated in full-frame mode. Given the optical brightness of our target ($V = 6.21$), the thick optical filter was used to reject optical and UV photons.
The EPIC data were processed with the \texttt{emchain} and \texttt{epchain} tasks from the Science Analysis Software (SAS) package (version 15.0; Current Calibration files as of 2017 March 2) to extract the event lists for the MOS and pn cameras.
We then selected the good time intervals (GTI) defined when the soft-proton flare count rate in the $0.2-10\,$keV energy range on the full detector is lower than $0.009$ and $0.004\,\mathrm{count\,s^{-1}\,arcmin^{-2}}$ for pn and MOS, respectively.
The first observation was only slightly affected by soft-proton flares (less than 20\% of the observational time), whereas the second observation had more than 75\% of contaminated time ranges.
We selected the X-ray events by keeping only the single and double events ($\mathrm{\texttt{PATTERN}\leq4}$) for the pn camera and the single, double, triple, and quadruple events ($\mathrm{\texttt{PATTERN}\leq12}$) for the MOS cameras.
Finally, we rejected the dead columns and bad pixels using the bit masks \texttt{FLAG==0} and \texttt{\#XMMEA\_SM} for the pn and MOS cameras, respectively.
The events from \stars{} (src+bkg) were extracted over a $15\arcsec$-radius circular area centered on the optical position of \stars{} \citep{gaia16a}.
This region allows us to extract 75 and 70\% of the flux at $1.5\,$keV on-axis for the pn and MOS cameras, respectively.
The contribution of the background (bkg) was estimated by extracting events from an approximately $3\arcmin \times 3\arcmin$ area located on the same CCD at about $4\arcmin$ north of \stars{}.
The X-ray point sources detected in the bkg region using the SAS task \texttt{edetect\_chain} were filtered out.

We also analyzed the Reflection Grating Spectrometer \citep[RGS;][]{denHerder} data.
We selected GTIs defined when the soft-proton flare count rate on the 9th CCD (on-axis) is lower than $0.1\,\mathrm{count\,s^{-1}}$.
We extracted the RGS spectra and the corresponding response files for the first and second order using the \texttt{rgsspectrum} and \texttt{rgsrmfgen} tasks, respectively.
Unfortunately, the second order spectra do not contain enough counts to be useful.
We thus only worked with the first order spectra of the RGS1 and RGS2 spectrometers.

\begin{table*}
\caption{Journal of the {\it XMM-Newton} observations of \stars{}.\label{journal}}
\begin{center}
\begin{tabular}{@{}ccccccccc@{}}
\hline
\hline
  Rev. & Date$\,$\tablefootmark{a} & $\phi\,$\tablefootmark{b} & Position angle$\,$\tablefootmark{c} & $d/a\,$\tablefootmark{d} & Exposure$\,$\tablefootmark{e} & pn$\,$\tablefootmark{f} & MOS1$\,$\tablefootmark{f} & MOS2$\,$\tablefootmark{f} \\
   & JD-2450000 & & ($^\circ$) & & (ks) & ($\mathrm{cts\ s^{-1}}$) & ($\mathrm{cts\ s^{-1}}$) & ($\mathrm{cts\ s^{-1}}$) \\
\hline
2713 & 6932.416 & $0.40\pm0.03$ & 123 & 1.094 & 26.8 & $0.116 \pm 0.002$ &  $0.054 \pm 0.001$ & $0.048 \pm 0.001$ \\
3078 & 7660.919 & $0.75\pm0.03$ & 228 & 1.012 & $\ \ $7.4 & $0.161 \pm 0.004$ & $0.030 \pm 0.002$ & $0.047 \pm 0.002$ \\
\hline
\end{tabular}
\tablefoot{
\tablefoottext{a} {Julian day at mid-exposure.}
\tablefoottext{b} {Orbital phase of the binary computed from the new orbital solution given in the bottom section of Table~\ref{solorb}.}
\tablefoottext{c} {The position angle is zero when the primary star is in front of the secondary.}
\tablefoottext{d} {Distance between the stars normalized by the semimajor axis computed from the new orbital solution.}
\tablefoottext{e} {Effective exposure time of the EPIC/pn camera after discarding soft-proton flares.}
\tablefoottext{f} {Net count rates for the three EPIC instruments over the $0.2-10\,$keV energy band.}
}
\end{center}
\end{table*}

\section{Analysis of the optical spectra}
\subsection{Orbital solution \label{RVsol}}
\label{rvsol}
\begin{figure}[h]
\begin{center}
\resizebox{8cm}{!}{\includegraphics{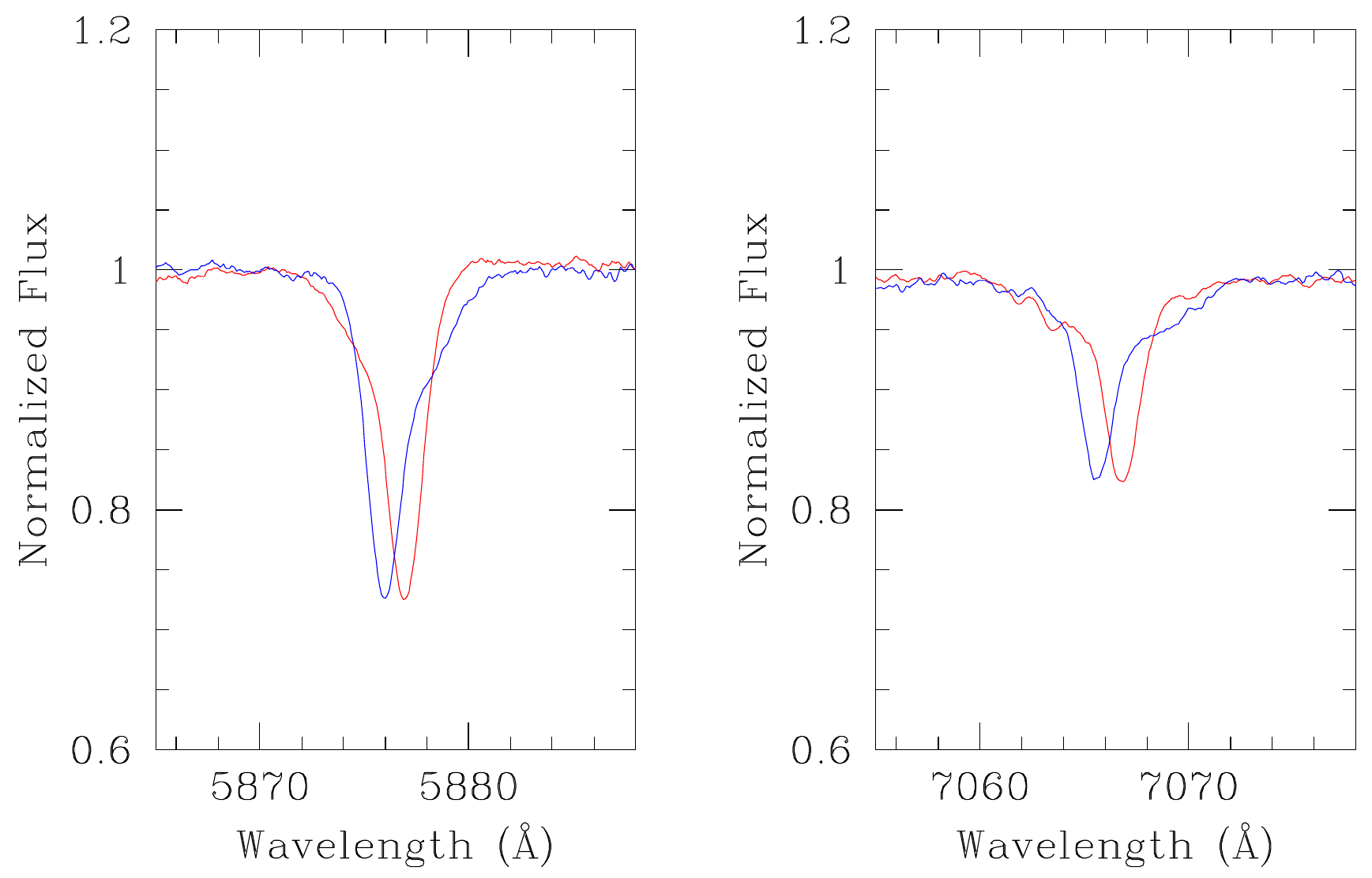}}
\end{center}  
\caption{He\,{\sc i} $\lambda$\,5876 (left) and He\,{\sc i} $\lambda$\,7065 (right) lines in the spectrum of HD\,54662 as observed with TIGRE+HEROS on HJD\,2456946.945 (2014, red) and 2457829.583 (2017, blue). Both lines reveal a clear asymmetry that reverses between the two epochs.\label{HeIprofiles}}
\end{figure}

Comparing the TIGRE observations taken at different epochs, we see a clear reversal in the asymmetry of the profile of the He\,{\sc i} $\lambda$\,5876 and He\,{\sc i} $\lambda$\,7065 lines between the spectra from 2014 and those of 2017 (see Fig.~\ref{HeIprofiles}). 
This behavior reflects the SB2 signature already pointed out by \citet{Boyajian}. 
 
As a first attempt to measure the RVs, we fitted two Gaussians to the observed He\,{\sc i} $\lambda$\,5876 and He\,{\sc i} $\lambda$\,7065 lines. 
While this process works very well for the sharp and prominent lines of the primary, the results are far more uncertain for the weaker, and significantly broader, secondary lines. 
We thus tried to improve the RVs using our spectral disentangling code based on the method outlined by \citet{GL}. 
The individual spectra were reconstructed in an iterative manner by averaging the observed spectra shifted into the frame of reference of one binary component after subtracting the current best approximation of the spectrum of the companion shifted to its current estimated RV. 
Improved estimates of the RVs of the stars were derived by cross-correlating a synthetic TLUSTY spectrum \citep{LH} with residual spectra obtained after subtracting the spectrum of the companion. 
Following \citet{VD}, the new estimates of the RVs were determined by fitting a parabola to the peak correlation function. 
For the synthetic TLUSTY spectra, we assumed effective temperatures of 37\,500\,K and $\log{g}=4.0$ for both stars. 
Initially, the synthetic spectral templates for both stars were rotationally broadened to $v\,\sin{i} \simeq 75$\,km\,s$^{-1}$ \citep[appropriate for the combined binary spectrum; see][]{Howarth}. 
From the reconstructed secondary spectra it became clear though that the spectral lines of the secondary are significantly broader. 
In a second attempt, we thus used $v\,\sin{i} \simeq 200$\,km\,s$^{-1}$ for the secondary template. 

\begin{figure}[h]
\begin{center}
\resizebox{8cm}{!}{\includegraphics{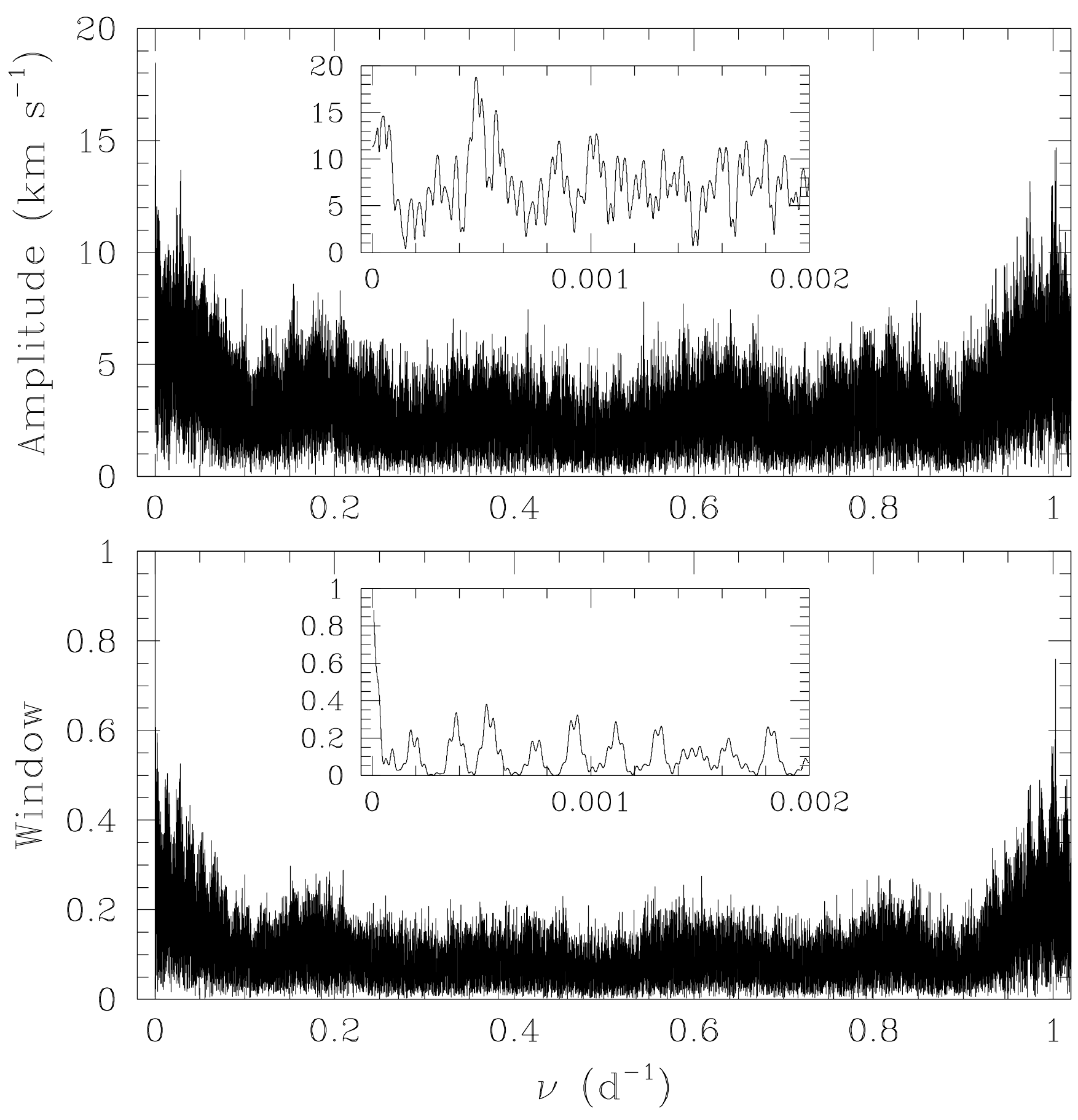}}
\end{center}  
\caption{Periodogram and spectral window of the primary RVs of \stars{} over 1\,d$^{-1}$.
The inner panels zoom on the region near the highest peak found at $4.752 \times 10^{-4}$\,d$^{-1}$.\label{fourier}}
\end{figure}

We performed a total of 55 iterations to disentangle the spectra. 
The resulting RVs are quoted in Table~\ref{journalopt}. 
Combining our new primary RVs with the literature data from \citet{Boyajian}, including the measurements from \citet{Plaskett}, \citet{Fullerton}, and the data from \citet{Garmany} and \citet{SL}, we obtained 106 measurements of the RVs of the primary spread  over 95 years. 
We then searched for periodicities using the Fourier method for uneven sampling of \citet{HMM}, modified by \citet{Gosset} and the trial period method of \citet{LK}. 
The Fourier method yields the highest peak at a frequency of $(4.752 \pm 0.029) \times 10^{-4}$\,d$^{-1}$ ($P = (2104\pm 13)$\,days, see Fig.~\ref{fourier}). 
The trial period method yields an estimate of the period of $2114\pm 13$\,days, which is consistent with the highest peak in the Fourier periodogram. 
In both cases, the error on the frequency was estimated conservatively as 0.1 times the natural width of the peak in the periodogram.
Our best estimate of the orbital period is thus much longer than the previous estimates of \citet{SL} and \citet{Boyajian}, but agrees very well with the values quoted by \citet{Sota} and \citet{LeBouquin}.

\begin{table}
\caption{Orbital solutions of \stars{}.}
\begin{center}
\begin{tabular}{c c c}        
\hline
\multicolumn{3}{c}{Full set of primary RVs}\\   
\hline
Element & \multicolumn{2}{c}{Value}\\
        & $e \neq 0$ & $e = 0$ \\
\hline
$P$ (days)          & \multicolumn{2}{c}{$2103.4 \pm 3.3$} \\
$v_0$ (km\,s$^{-1}$) & $42.8 \pm 0.3$ & $42.7 \pm 0.2$\\
$e$                 & $\leq 0.02$ & 0.0 (assumed)\\
$K_1$ (km\,s$^{-1}$) & $19.7 \pm 0.4$ & $19.8 \pm 0.3$ \\
$\omega$ ($^{\circ}$) & $193.3 \pm 64.5$ & \\
$T_0$\,\tablefootmark{a} (HJD-2450000) & $5787 \pm 376$ & $6233.6 \pm 5.2$\\
$a\,\sin{i}$ ($\rsun{}$) & $817 \pm 15$ & $821 \pm 13$ \\
$f(m)$ ($\msun{}$) & $1.66 \pm 0.29$ & $1.68 \pm 0.08$ \\
rms (km\,s$^{-1}$) & $2.30$ & $2.15$ \\
\hline
\end{tabular}
\begin{tabular}{c c c}        
\hline
\multicolumn{3}{c}{Primary and secondary RVs}\\   
\hline
Element & \multicolumn{2}{c}{Value}\\
        & Primary & Secondary \\
\hline
$P$ (days)          & \multicolumn{2}{c}{$2103.4$ (fixed)} \\
$m2/m1$             & \multicolumn{2}{c}{$0.843 \pm 0.032$} \\
$v_0$ (km\,s$^{-1}$) & $42.5 \pm 1.2$ & $35.9 \pm 1.2$\\
$e$                 & \multicolumn{2}{c}{$0.11 \pm 0.02$}\\
$K$ (km\,s$^{-1}$) & $19.3 \pm 0.6$ & $22.9 \pm 0.7$ \\
$\omega$ ($^{\circ}$) & \multicolumn{2}{c}{$240.9 \pm 10.9$}  \\
$T_0$\,\tablefootmark{a} (HJD-2450000) & \multicolumn{2}{c}{$6084 \pm 62$}\\
$a\,\sin{i}$ ($\rsun{}$) & $797 \pm 24$ & $945 \pm 28$ \\
$m\,\sin^3{i}$ ($\msun{}$) & $8.72 \pm 0.64$ & $7.35 \pm 0.53$ \\
rms (km\,s$^{-1}$) & $1.72$ & $2.52$ \\
\hline
\end{tabular}
\end{center}
\tablefoot{
\tablefoottext{a} {For eccentric orbital solutions, $T_0$ corresponds to the time of periastron passage, whereas it stands for the time of conjunction with the primary passing in front in the case of a circular orbital solution.}
}
\end{table}

\begin{figure}[h]
\begin{center}
\resizebox{8cm}{!}{\includegraphics{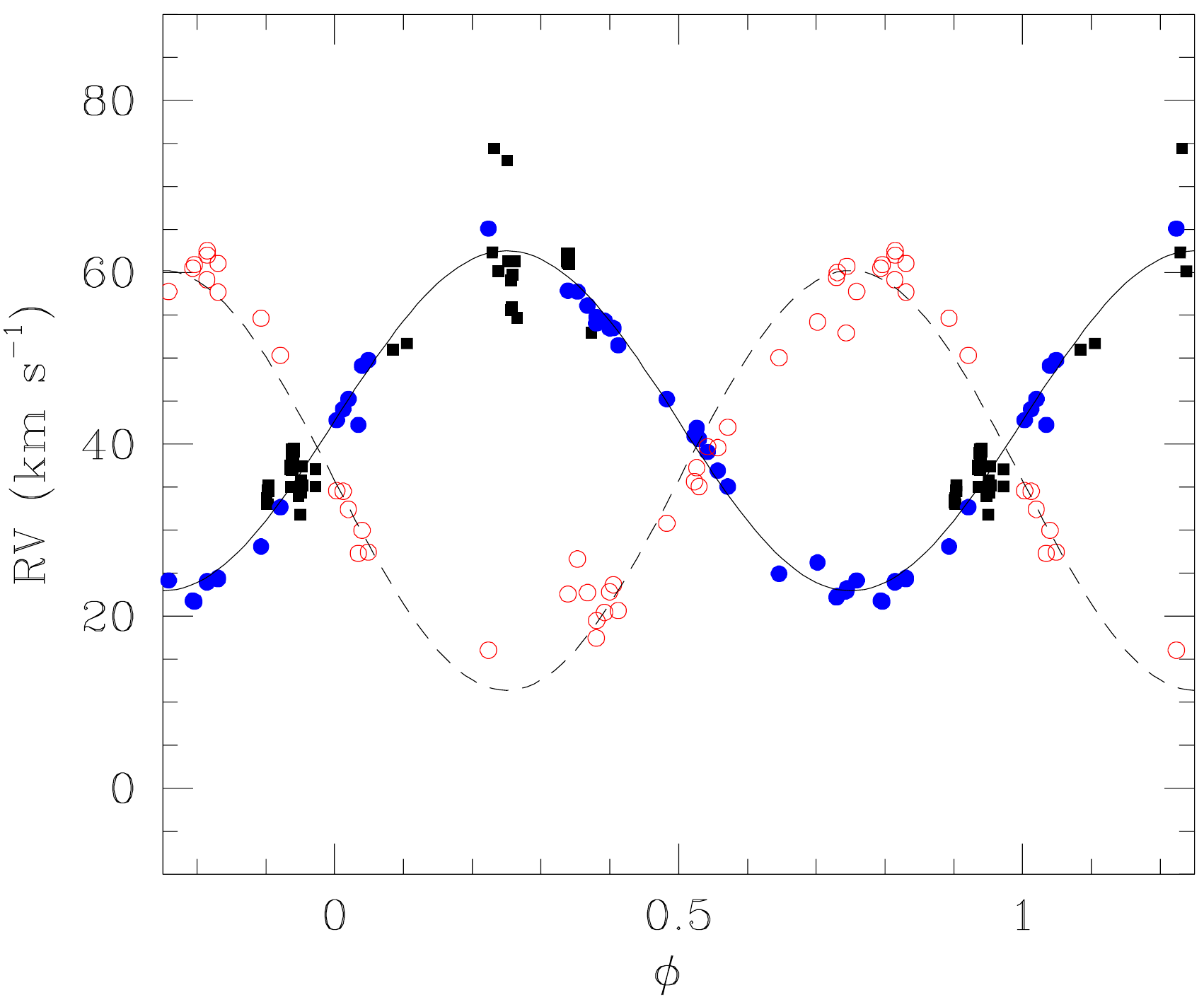}}
\resizebox{8cm}{!}{\includegraphics{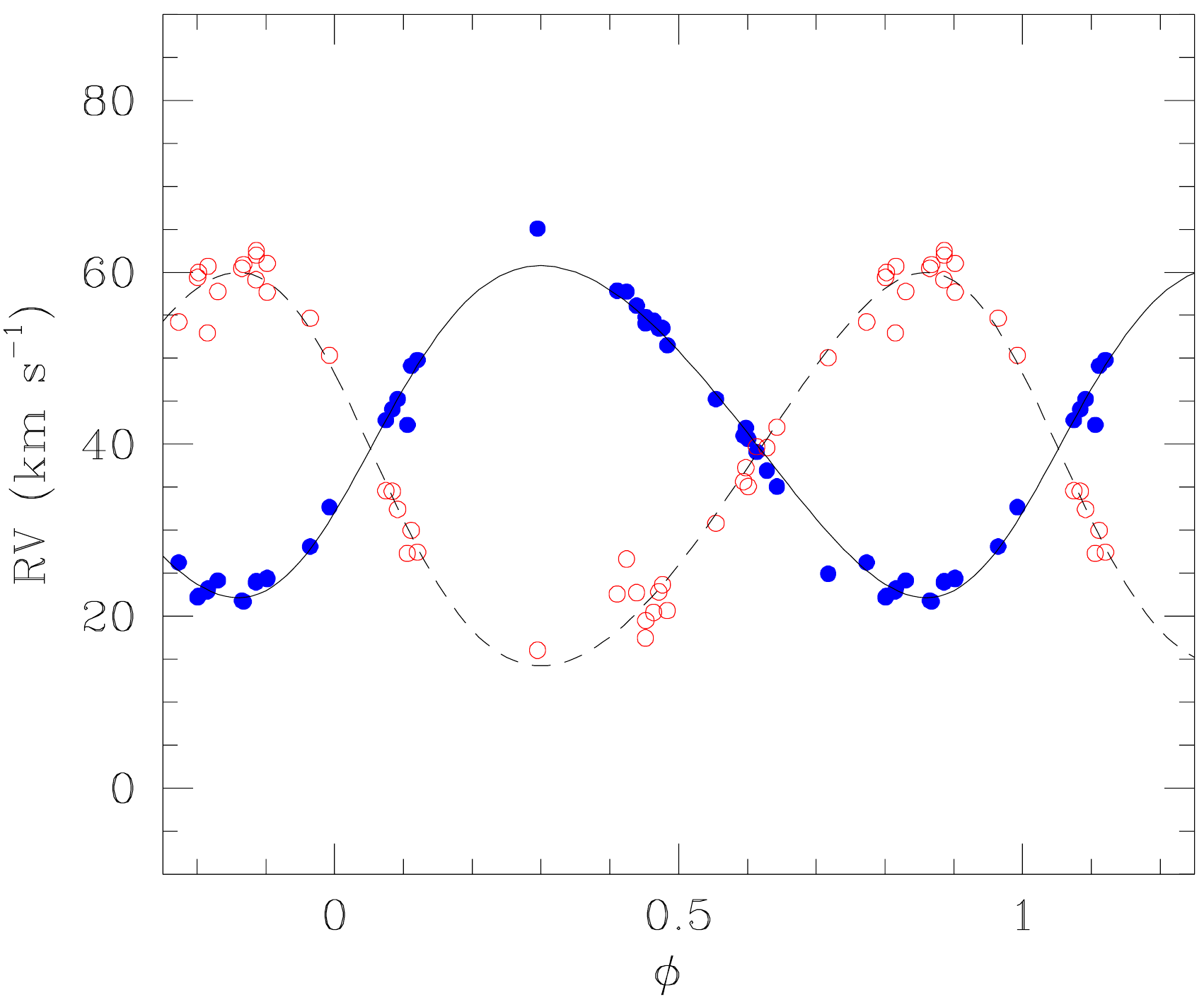}}
\end{center}  
\caption{Orbital solution of \stars{}. In the top panel, an SB1 solution was computed from the full set of primary RVs from the literature (filled squares) and newly determined values (filled circles) assuming $e=0$. 
The mass-ratio determined from the new data was then used to compute the secondary RV curve that best matches the secondary RV data (open circles). 
The bottom panel shows an SB2 orbital solution (which has $e = 0.11$) computed using only the new RV data.\label{solorb}}
\end{figure}

With this estimate of the orbital period as input and using the full set of 106 primary RVs, we then computed an SB1 orbital solution by means of the Li\`ege Orbital Solution Package \citep[LOSP;][]{SGR}\footnote{LOSP is an improved version of the code originally proposed by \citet{WHS}, maintained by H.~Sana and available at \href{http://www.stsci.edu/~hsana/losp.html}{http://www.stsci.edu/~hsana/losp.html}}. 
The resulting orbital period refined by the LOSP code is $2103.4 \pm 3.3$\,days.
This value is consistent with the periods found by the Fourier and the trial method but the estimate by LOSP is better constrained since this method assigns a weight to the data as a function of their error bars.
Assuming an eccentric orbit, we found that the best-fit value of $e$ is consistent with 0.
As a next step, we thus tested a circular orbit. 
The best orbital parameters are given in Table~\ref{solorb} and the SB1 orbital solution is shown with the solid line in the top panel of Fig.~\ref{solorb}.
To check whether the nonzero eccentricity is significant, we applied the method of \citet{lucy71}.
Using our primary RVs, we computed the sum of the square of the deviations from the orbital solution: for the circular and elliptical solution, we have $R_\mathrm{c}=589.1\,(\mathrm{km\,s^{-1}})^2$ and $R_\mathrm{e}=568.9\,(\mathrm{km\,s^{-1}})^2$, respectively.
We then computed the probability that the orbit is actually circular as
\begin{equation}
   p=\left(1+\frac{(N-M)(R_\mathrm{c}-R_\mathrm{e})}{2R_\mathrm{e}\beta}\right)^{-\beta}
,\end{equation}
where $N=106$ is the number of data points, $M$ the number of orbital parameters, and $\beta=(N-M)/2$.
We thus have $p=0.18$.
Adopting a 5\% confidence level, the nonzero eccentricity is not significant.

Using the secondary RVs obtained via spectral disentangling, we then estimated the mass ratio $m_2/m_1 = 0.84 \pm 0.03$. The corresponding orbital solution for the secondary is shown by the dashed line in the top  panel of Fig.~\ref{solorb}. Our estimate of the mass ratio is significantly larger than the value $0.39 \pm 0.06$ quoted by \citet{Boyajian}. This discrepancy most likely stems from the difficulty in measuring the weak and broad signature of the secondary star on the spectra of \citet{Boyajian}. Moreover, their data did not sample the full RV excursion of the primary, leading to a lower value of $K_1$. 

Since the older RVs of the primary star are subject to larger uncertainties, we also computed an SB2 solution applying LOSP on the 39 RVs listed in Table~\ref{journalopt} and adopting an orbital period of 2103.4\,days as obtained for the full set of primary RVs. The  best orbital parameters are listed in the bottom section of Table~\ref{solorb}, and the solution is illustrated in the bottom panel of Fig.~\ref{solorb}.
The application of the test of \citet{lucy71} leads to a probability that the orbit is circular of $p=2.17\times 10^{-8}$ proving that the eccentricity of 0.11 is not a spurious value.

The corresponding minimum masses of the primary and secondary are $m_1\,\sin^3{i} = (8.72 \pm 0.64)$\,M$_{\odot}$ and $m_2\,\sin^3{i} = (7.35 \pm 0.53)$\,M$_{\odot}$. 
Adopting an inclination of $i = 74.87^{\circ}$ as inferred by \citet{LeBouquin}, our orbital solution implies absolute masses of $m_1 = 9.7 \pm 0.7$\,M$_{\odot}$ and $m_2 = 8.2 \pm 0.6$\,M$_{\odot}$ for the primary and secondary, respectively. 
These values are surprisingly low for O-type stars \citep{martins05,weidner10}. 
Making the dynamic masses of the stars consistent with typical masses of stars of their spectral type would require a much lower inclination around $38^{\circ}$. 

A similar discrepancy between the masses inferred by combining the RV solution with the astrometric solution and typical masses was already found for another long-period massive binary (9~Sgr; \citealt{9Sgr}). 
Assuming, we are dealing with genuine Population I O-type stars, the problem could be due to a bias in either the astrometric or RV solution. 
Concerning the astrometric solution, we note that the fact that the orbit is nearly circular should in principle lead to a rather robust determination of the inclination as it only depends on the shape of the orbit projected on the sky. 
Still, comparing the semimajor axis projected on the plane of the sky as measured by \citet{LeBouquin} with our value of $a\,\sin{i}$, we find that an inclination of $38^{\circ}$ leads to a distance of 1.27\,kpc which is in good agreement with the distance of 1.2\,kpc inferred by \citet{kaltcheva00}. 
Adopting instead the orbital inclination estimated by \citet{LeBouquin} would imply a much smaller distance of about 800\,pc. 
As to the possibility of a bias in the RV solution, we note that the RVs of the sharp-lined primary star are well determined and its orbital solution is thus robust. 
The situation is more complex for the broad-lined secondary star for which establishing precise RVs is more difficult. 
Given the mass function established from the primary RVs (see Table\,3), a mass ratio $m_2/m_1$ of 0.530 would be required to obtain a primary mass of 28\,M$_{\odot}$, typical for an O6.5\,V star \citep{martins05}, for an inclination of $74.87^{\circ}$. 
This mass ratio is outside the error bars of our observational determination and would imply a secondary mass of $14.8\,\msun{}$, which is more typical of a B0 star than of an O6.5. 
Finally, there is still a possibility  that we could be dealing with a pair of subdwarf O stars. 
The latter objects form a rather heterogeneous group \citep{heber09}, but are mostly thought to result from the evolution of stars with masses well below the minimum masses we have inferred for the components of \stars{}. 
Moreover, the properties of the components ($\log{g}$, chemical abundances) that we have determined with CMFGEN do not cope with the sdO scenario either. 

As already mentioned by several authors, low eccentricities are surprising for such long-period binary systems (e.g., \citealt{neilson15,LeBouquin}).
The contribution of the primary component to the circularization timescale may be computed as \citep{zahn77,KK}
\begin{eqnarray}
 t_\mathrm{circ,1}^{-1} & \sim & 6.59 \times 10^{-3}\,\left(\frac{m_1\,R_{\odot}^3}{M_{\odot}\,R_1^3}\right)^{1/2}\,q\,(1 + q)^{11/6}\,E_{\rm tidal}\,\left(\frac{R_1}{a}\right)^{21/2} \nonumber \\
 & & F_{212}\,A_\mathrm{z}\,\mathrm{s}^{-1}
,\end{eqnarray}
where $E_{\rm tidal}$ is the tidal coefficient lower than $10^{-5.5}$ for a $30$\,M$_{\odot}$\,star on the main sequence \citep{KK}, $q = m_2/m_1$, and $R_1$ the radius of the primary. 
The terms $F_{212}(e) = \frac{1}{2\,\pi}\,\int_0^{2\,\pi} (1 - e\,\cos{E})^{-19/2}\,dE $ and $A_\mathrm{z}=2$ were introduced by \citet{KK} to account for the increase of the efficiency of the circularization mechanism with the eccentricity. 
For $e = 0.11$, we obtain $F_{212} =1.34$.

The contribution of the secondary star can be computed likewise by inverting the indices 1 and 2. The circularization timescale is then obtained from 
\begin{equation}
t_\mathrm{circ}^{-1} = t_\mathrm{circ,1}^{-1} + t_\mathrm{circ,2}^{-1}.
\end{equation}
Assuming that both stars have a radius of about $10\,\rsun{}$ (typical for stars with such spectral types) and considering the orbital parameters deduced from the SB2 solution, we obtain
\begin{equation}
t_\mathrm{circ} \sim  2.1\times 10^{8}\,E_{\rm tidal}^{-1}\,(\sin{i})^{-9}\,\mathrm{yr.}
\end{equation}
Since $E_{\rm tidal} \leq 10^{-5.5}$ and $\sin{i} \leq 1$, the circularization timescale is much larger than the lifetime of an O-type star on the main sequence (which is on the order of $5\times10^6\,$yr).
The \stars{} system is thus an atypical case of very-long period nearly circular orbit.

\subsection{Reconstructed spectra}
\begin{figure*}[htb]
\resizebox{!}{7.1cm}{\includegraphics{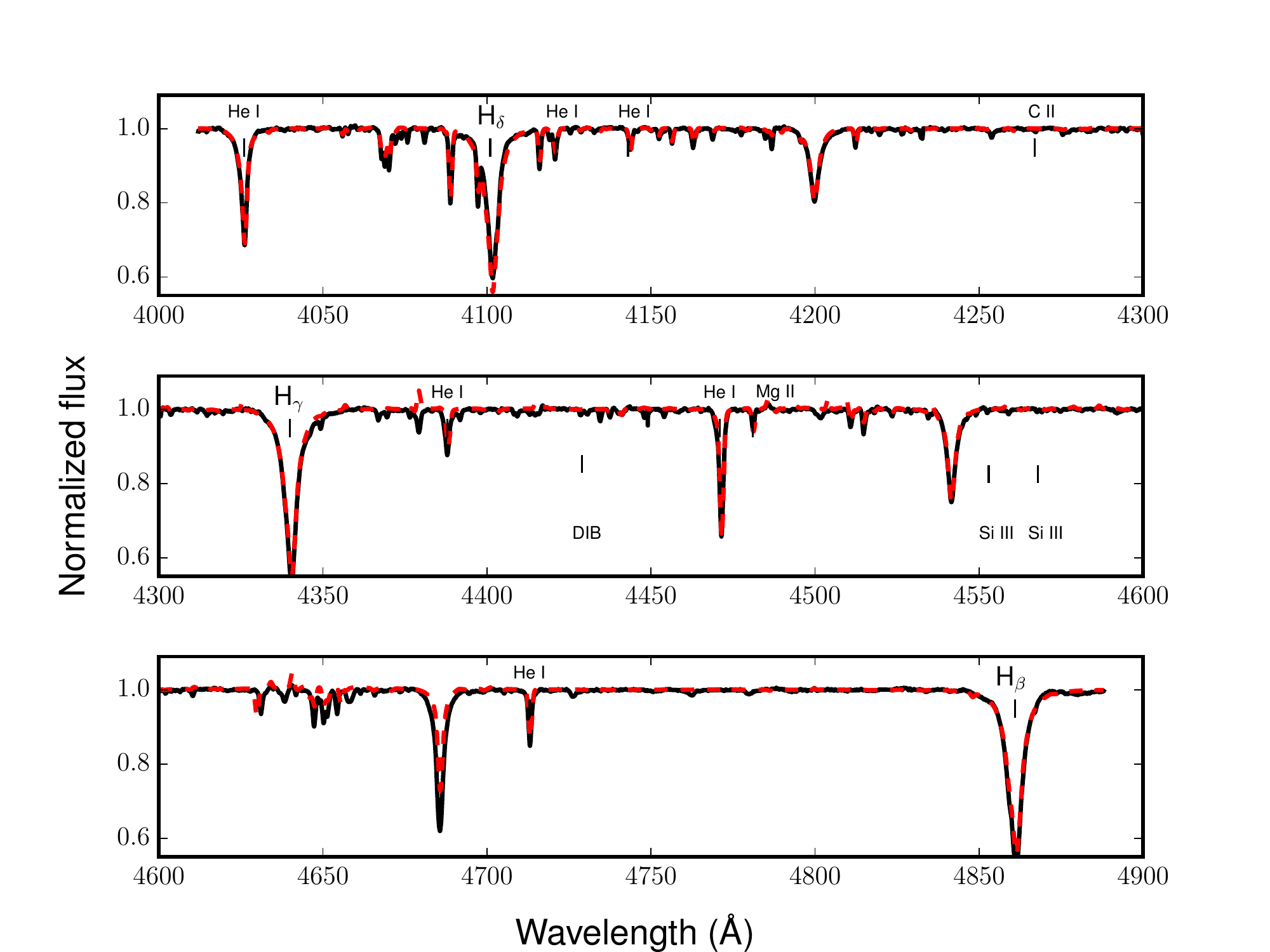}}
\resizebox{!}{7.1cm}{\includegraphics{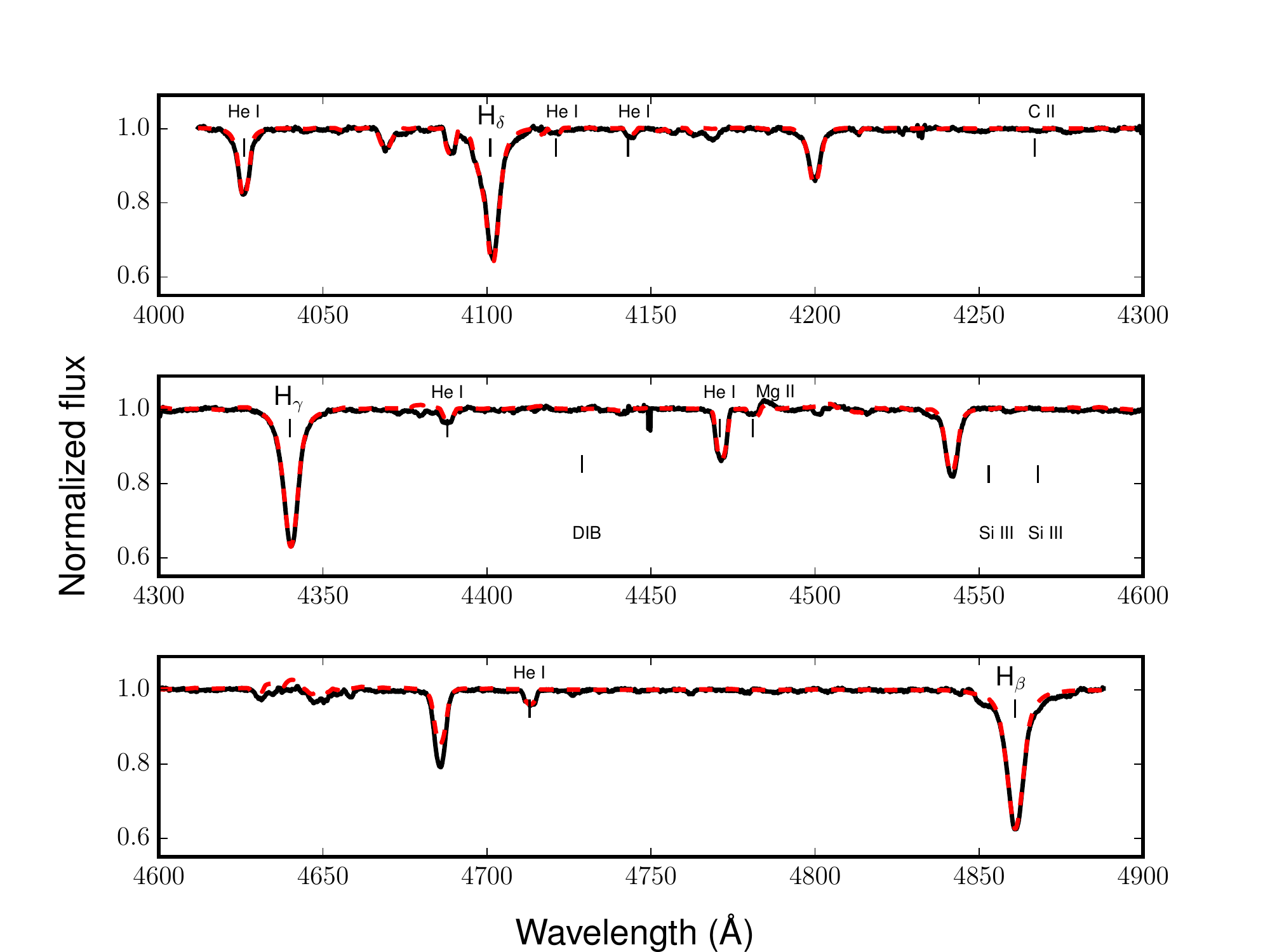}}
\caption{Reconstructed spectra of the primary (left) and secondary (right) star of \stars{}. 
The black solid lines illustrate the spectra obtained through spectral disentangling whilst the red dashed lines correspond to the best-fit CMFGEN models (see text). 
The reconstructed spectra were normalized to their continuum assuming the primary is twice as bright as the secondary in the optical domain.} \label{disent}
\end{figure*}

Disentangled spectra of the primary and secondary star are shown in Fig.~\ref{disent}, along with the best-fit CMFGEN  spectra (see Sect.\,\ref{cmfgen}). 

The combined spectrum of \stars{} was classified as O7\,Vz var? by \citet{Sota}.
The z tag indicates that the He\,{\sc ii} $\lambda$\,4686 line is stronger than other He\,{\sc ii} lines \citep{walborn97,arias14}. 
To perform the spectral classification of the binary components, we measured the equivalent widths (EW) of the He\,{\sc i} $\lambda$\,4471 and He\,{\sc ii} $\lambda$\,4542 lines in the reconstructed primary and secondary spectra. 
We inferred O6.5 spectral types for both stars \citep{CA, CF}. 
From the ratio of the primary and secondary EWs of the same lines, we then concluded that the primary star should be about two times brighter than the secondary in the optical domain, corresponding to a magnitude difference of about $0.75$ (see also Sect.\,\ref{cmfgen}). 
This value is larger than the magnitude difference of $0.23 \pm 0.04$ in the $H$ band inferred from interferometry \citep{SMASH}.

We used the disentangled spectra to determine the projected rotational velocities $v\,sin{i}$ of the components of \stars{}. 
For this purpose, we applied the Fourier transform method \citep{Gray,Simon-Diaz} to the profiles of the He\,{\sc i} $\lambda$\,4713 line of the primary and the He\,{\sc i} $\lambda\lambda$\,4144, 4388, 4471, 4713, He\,{\sc ii} $\lambda\lambda$\,4200, and 4542, Mg\,{\sc ii} $\lambda$\,4481 lines of the secondary.
For the primary, we obtained $v\,\sin{i}$ of 43\,km\,s$^{-1}$ and we further estimated that a macroturbulent velocity of 35\,km\,s$^{-1}$ is needed to match the observed line profile with synthetic spectra (see Sect.\,\ref{cmfgen}). 
For the secondary, we obtained $v\,\sin{i}$ of $158 \pm 5$\,km\,s$^{-1}$. 

\subsection{Model atmosphere fitting \label{cmfgen}}
We used the non-LTE atmosphere code CMFGEN \citep{hillier98} to constrain the chemical and physical properties of both stars of \stars{}.
The CMFGEN code is designed to model O-type stars, Wolf-Rayet stars, luminous blue variables, and supernovae by the resolution of the radiative transfer and statistical equilibrium equations in the co-moving frame.
This code also accounts for the effect of line blanketing on the energy distribution and uses a $\beta$ velocity law for the stellar winds.
The chemical species and their ions included in our CMFGEN models are H, He, C, N, O, Si, S, Ne, Mg, Al, Ar, Fe, and Ni.
We computed a grid of CMFGEN models with $T_\mathrm{eff}$ between 25000\,K and 46000\,K ($\Delta T_\mathrm{eff}=1000\,$K) and $\log{} g$ between 3.0 and 4.3 ($\Delta \log{} g = 0.1$). 
The wind prescription was taken from \citet{vink01} and we used standard luminosities \citep{martins05} to compute those models. 
At first, the surface abundances were fixed to the solar values of \citet{grevesse07}. 
We then proceeded to a determination of the $\chi^2$ between disentangled spectra and synthetic models computed on diagnostic lines (\ion{He}{I}/\ion{He}{II} for the effective temperature and the wings of the Balmer lines for $\log{} g$).
The microturbulence was taken to vary linearly with the wind velocity starting from $10\,\mathrm{km\,s^{-1}}$ in the photosphere to a maximum of $0.1\,v_\infty$. 
The hydrodynamical structure of the stellar atmosphere is given as an input parameter of the code.

We determined effective temperatures of $T_\mathrm{eff,1}=37700\pm1000\,$K and $T_\mathrm{eff,2}=37500\pm1000\,$K for the primary and secondary star, respectively.
We also determined a $\log{} g$ of $3.96\pm0.1$ and $3.81\pm0.1$ for the primary and secondary star, respectively.
We stress that the errors on the surface gravities correspond to the errors of the fit only and do not account for the uncertainties on the reconstruction of the wings of the Balmer lines. 
Indeed,  it is difficult to reproduce broad features such as the Balmer lines via spectral disentangling of binary systems, and the real errors on log g are thus likely higher.

The He lines were well matched with a number abundance of He/H=0.1 (i.e., close to solar).
The abundances of nitrogen and carbon were determined using the method of \citet[][see also \citealt{mahy17} for an application to a binary system]{martins15}.
For the carbon abundances, we mainly focused on the \ion{C}{III} $\lambda\lambda$4068-70 lines, since the \ion{C}{III} lines around 4650$\,\AA{}$ are affected by the wind parameters (see \citealt{martins12}). 
For the determination of the nitrogen surface abundances, we focused on the triplet around 4515$\,\AA{}$.
In this way, we found a significant overabundance of N in both stars, especially for the primary (see Table~\ref{chemi_table}).
\begin{table}
\centering
\caption[Chemical abundances of the components of \stars{}]{Chemical abundances of the components of \stars{}.}
\label{chemi_table}
\resizebox{0.4\textwidth}{!}{
\begin{tabular}{@{}cccc@{}}
\hline
\hline
 & Primary & Secondary & Sun$\,$\tablefootmark{a} \\
\hline
C/H & $(2.6\pm0.3)\times 10^{-4}$ & $(2.3\pm0.2)\times 10^{-4}$ & $2.69\times 10^{-4}$ \\
N/H & $(2.3\pm0.3)\times 10^{-4}$ & $(9.5\pm1.5)\times 10^{-5}$ & $6.76\times 10^{-5}$ \\
\hline
\end{tabular}
}
\tablefoot{
\tablefoottext{a} {Taken from \citet{asplund09}.}
}
\end{table}

\section{Analysis of the X-ray spectra}
\label{xray_spectra}
To study the X-ray spectra of \stars{}, we used the src+bkg and the bkg event lists created in Sect.~\ref{xmm_obs}.
We constructed the EPIC spectra, ancillary files, and response matrices for the two observations using the SAS task \texttt{especget}.
The extracted spectra were then grouped using the SAS task \texttt{specgroup} starting at $0.2\,$keV with a minimum signal-to-noise ratio of 10 and 5 for the first and second observation, respectively. 
The first order RGS spectra were grouped with a minimum signal-to-noise ratio of 3. 
Because of the presence of a strong photospheric UV radiation field from the massive stars and, possibly, a high-density wind-shock region, the forbidden lines of He-like triplets could be depopulated in comparison to the intercombination lines \citep[][and references therein]{Porquet}. 
The number of counts in the RGS spectra is too low to observe the depopulation but we grouped together the resonant, intercombination, and forbidden lines of the \ion{O}{VII}, \ion{N}{VI,} and {IX} ions to avoid any bias from depopulation in the spectral fitting.

We fit the pn, MOS1, MOS2, RGS1, and RGS2 spectra of each observation simultaneously using an absorbed APEC model \citep{apec} reproducing the emission spectrum for collisionally ionized diffuse gas using the atomic emission line database AtomDB v.3.0.7.
The absorption of the X-ray emission is produced by two components: the hydrogen column density of the neutral ISM along the line of sight (created using \texttt{TBnew}; \citealt{wilms00} and the dust scattering model \texttt{dustscat}$\,$\footnote{Since \texttt{TBnew} uses the cross sections from \citet{verner96} and the ISM abundances of \citet{wilms00}, the column density derived from \texttt{dustscat} must be set to those derived from \texttt{TBnew} divided by 1.5 \citep{nowak12}.}; \citealt{predehl95}) and the column density of the stellar wind (modeled with the tabulated \texttt{wind} model; \citealt{naze04}).
The ISM hydrogen column density was fixed to $1.68 \times 10^{21}\ \mathrm{cm^{-2}}$ \citep{shull85,diplas94}.

\begin{figure}
\centering
\includegraphics[trim=0.cm 0.8cm 0.cm 1.5cm,clip,height=8.3cm,angle=0]{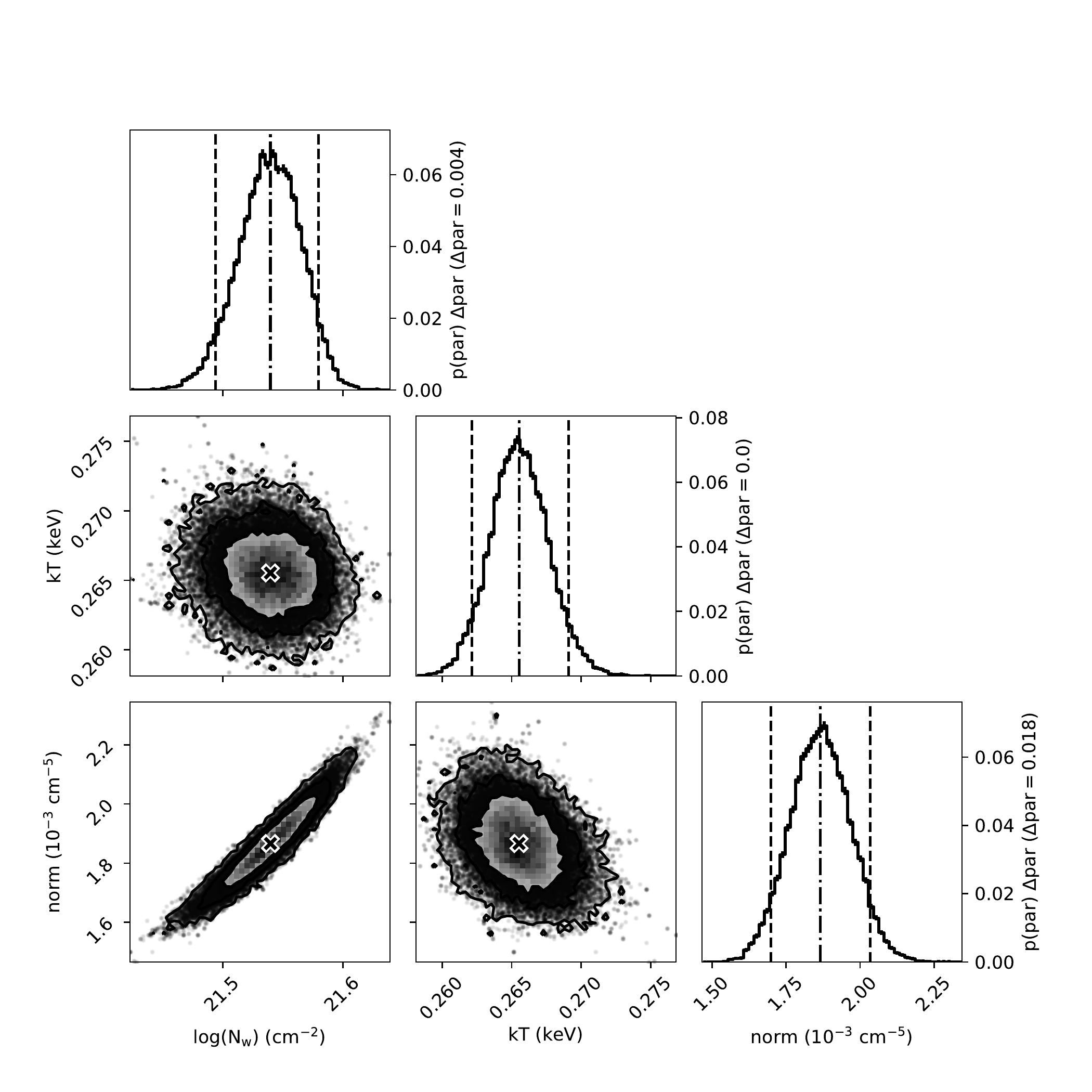}
\includegraphics[trim=0.cm 0.8cm 0.cm 1.5cm,clip,height=8.3cm,angle=0]{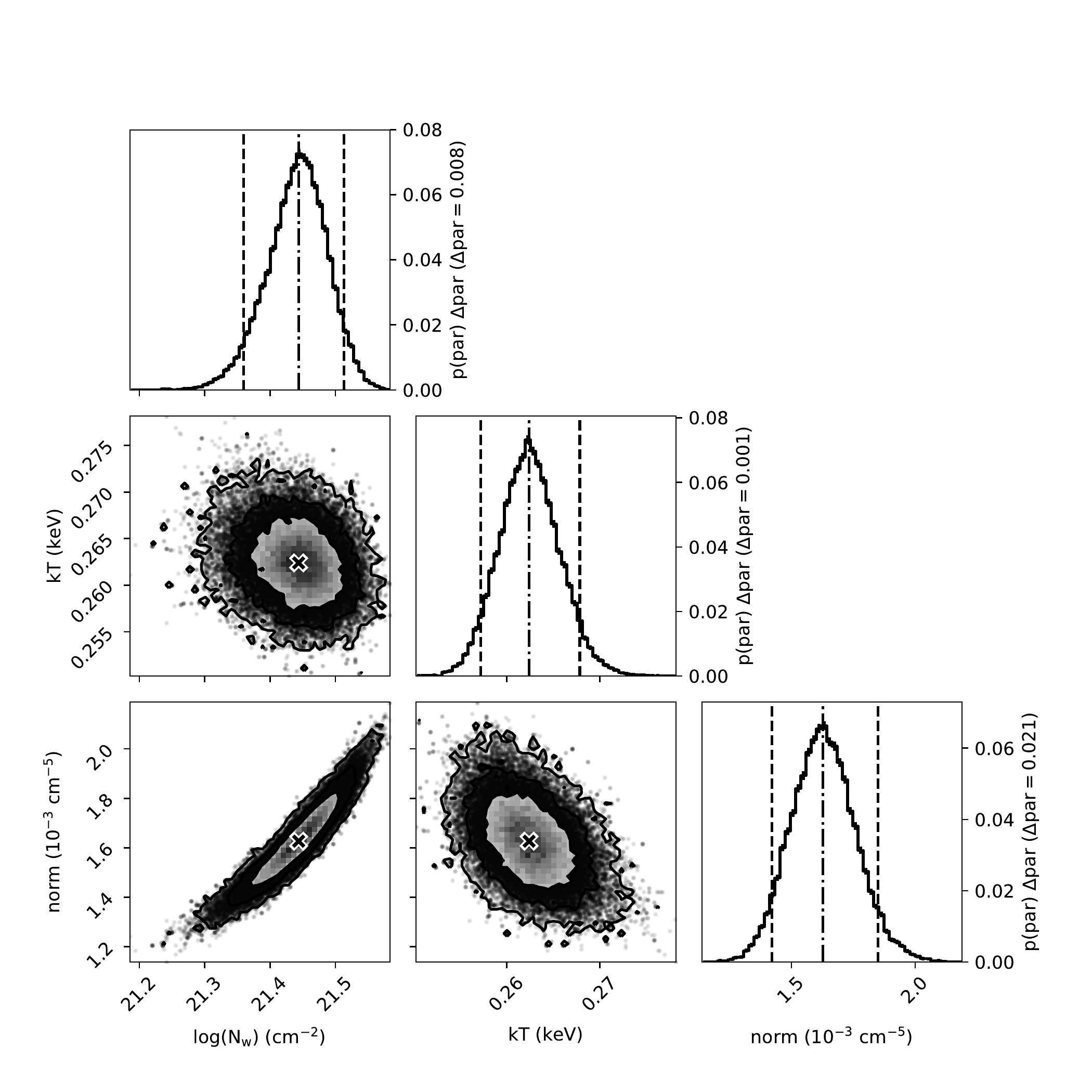}
\caption[Best-fit spectral parameters of \stars{} during the 2014 and 2016 observations]{Best-fit spectral parameters of \stars{} during 2014 (top) and 2016 (bottom) observations.
The diagonal plots are the marginal distribution of each parameter. 
The median values of the marginal distributions are the vertical dotted lines in diagonal plots and the crosses in other panels.
The vertical dashed lines define the 90\% confidence interval. 
The contours are 68\%, 90\%, and 99\% confidence levels.
}
\label{mcmc_fig}
\end{figure}

\begin{figure}
\centering
\includegraphics[trim=1.cm 1.cm 0.5cm 3.cm,clip,height=9cm,angle=-90]{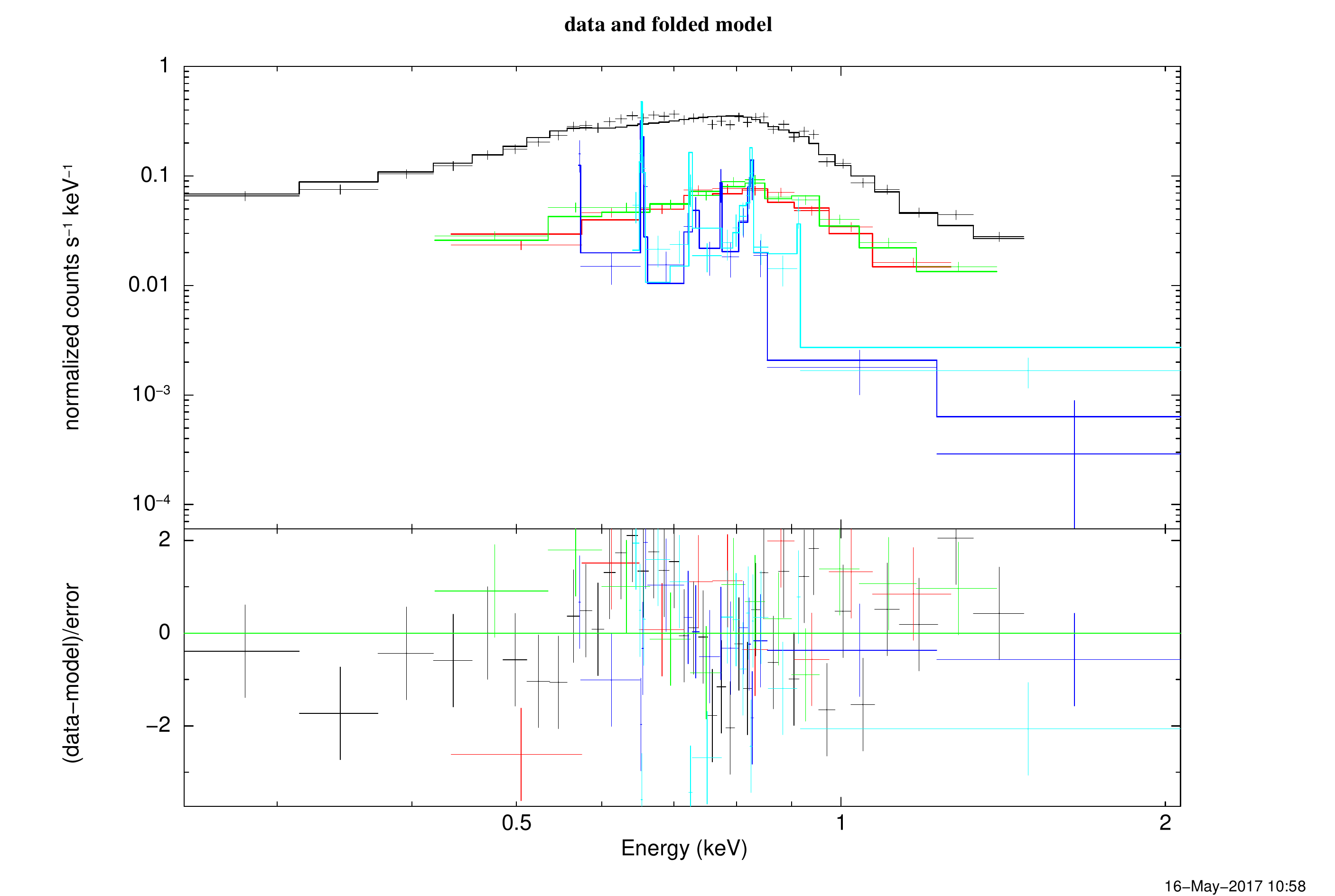}
\includegraphics[trim=1.cm 1.cm 0.5cm 3.cm,clip,height=9cm,angle=-90]{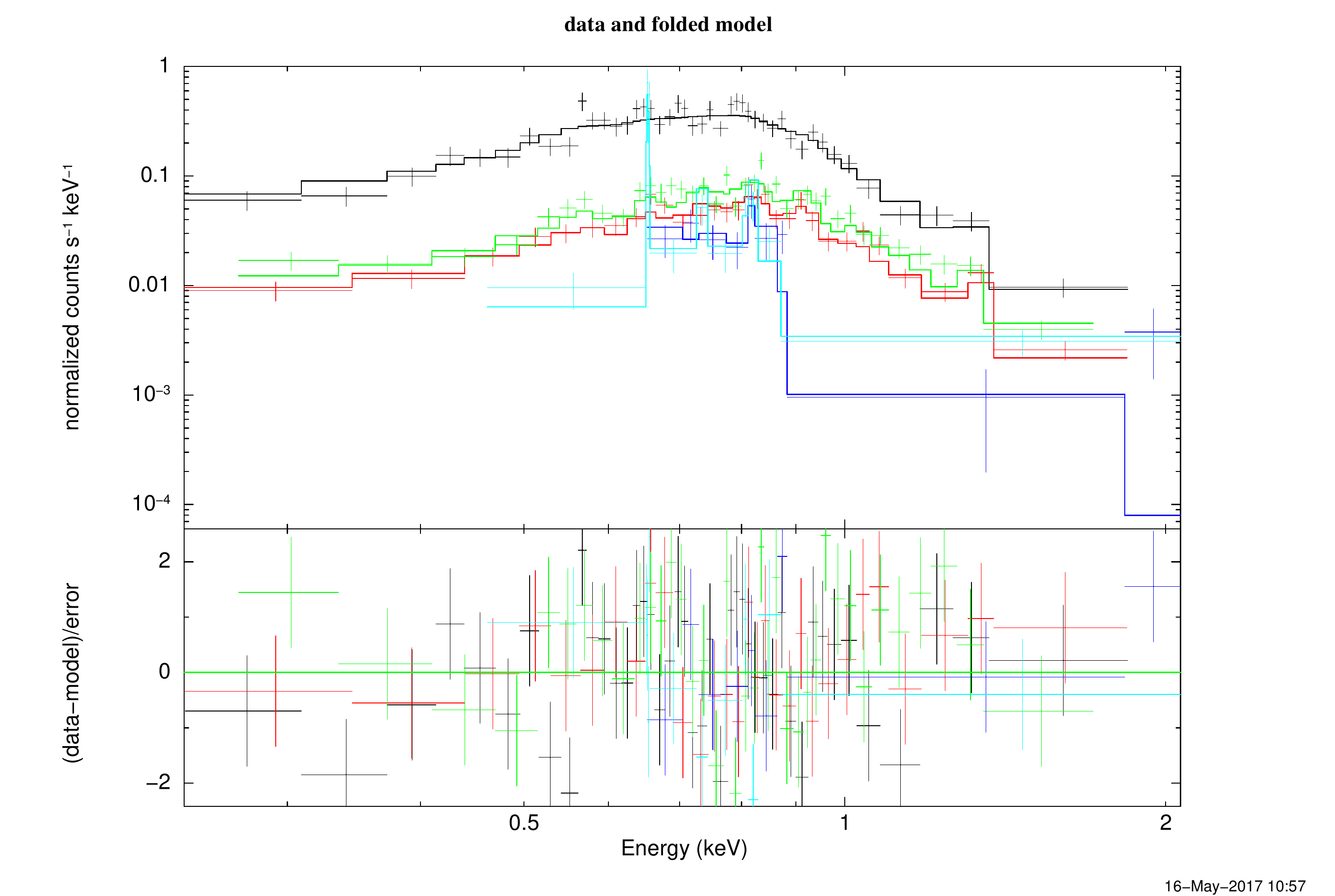}
\caption[X-ray spectra of \stars{} during the 2014 and 2016 observations]{X-ray spectra of \stars{} during the 2014 (top) and 2016 observations (bottom).
The black, green, red, light blue, and dark blue lines are the pn, MOS1, MOS2, RGS1 (first order), and RGS2 (first order) spectra, respectively.
The bottom panel in the two graphs shows the residuals.
}
\label{X_ray_spectra}
\end{figure}

We used the Markov Chain Monte Carlo (MCMC) algorithm to constrain the three parameters of the model: the column density of the cool stellar wind ($N_\mathrm{w}$), temperature of the X-ray plasma ($kT$), and normalization of the APEC model ($norm$).
The MCMC is an iterative method producing a set of walkers evolving simultaneously and converging toward the marginal distribution of each model parameters describing the observed spectrum.
The values of the model parameters taken by the walkers at each step are determined by the positions of the walkers at the previous step and the new set of parameters is accepted if the value of the likelihood function is lower than the previous value.
The MCMC algorithm used in this work (\texttt{XSPEC\_emcee}$\,$\footnote{\href{https://github.com/jeremysanders/xspec\_emcee}{https://github.com/jeremysanders/xspec\_emcee}}) fits the spectra using XSPEC (version 12.9.1a) and an extensible, pure Python implementation of the \citet{goodman10} affine invariant MCMC ensemble sampler (\texttt{emcee}$\,$\footnote{\href{http://dan.iel.fm/emcee/current/user/line}{http://dan.iel.fm/emcee/current/user/line}}; \citealt{foreman-mackey13}).
We \textit{a posteriori} checked the convergence of the parameters by computing the mean acceptance fraction whose good range (based on a huge number of simulations) is between 0.2 and 0.5 \citep{gelman96,foreman-mackey13}.
The mean acceptance fraction is around 0.66 for the two observations, which is a reliable value.
The best-fit parameter values are the median (i.e., 50th percentile) of the marginal distribution of each parameter (see diagonal plots of Fig.~\ref{mcmc_fig}). 
We also defined a 90\% confidence range for each parameter as the 5th and 95th percentile of their marginal distributions. 
These numbers are reported in Table~\ref{mcmc_table}. 
The corresponding best spectra are shown in Fig.~\ref{X_ray_spectra}.

One can observe that the X-ray spectra of the 2014 and 2016 observations are very soft with nearly all events having an energy lower than $2\,$keV unlike what is expected for wide colliding wind binaries \citep{stevens92}.
The X-ray emission of \stars{} is thus probably dominated by the intrinsic emission from the stars.

\begin{table}
\centering
\caption[Spectral properties of \stars{} observed by XMM-Newton]{Spectral properties of \stars{} observed by {\it XMM-Newton}.}
\label{mcmc_table}
\resizebox{0.49\textwidth}{!}{
\begin{tabular}{@{}cccccc@{}}
\hline
\hline
Rev. & $\log_{10}(N_\mathrm{w})$ & $kT$ & $norm$ & $F^\mathrm{unabs}_\mathrm{0.2-10keV}\,$\tablefootmark{a} & $\chi^2_\mathrm{red}\,$\tablefootmark{b} \\
 & ($\mathrm{cm^{-2}}$) & (keV) & ($10^{-3}\ \mathrm{cm^{-5}}$) & ($10^{-12}\ \mathrm{erg\,cm^{-2}\,s^{-1}}$) & \\
\hline
2713 & $21.54^{+0.04}_{-0.05}$ & $0.266^{+0.004}_{-0.003}$ & $1.9^{+0.2}_{-0.2}$ & $1.30^{+0.03}_{-0.03}$ & $154/94$ \\
3078 & $21.44^{+0.07}_{-0.08}$ & $0.262^{+0.005}_{-0.005}$ & $1.6^{+0.2}_{-0.2}$ & $1.34^{+0.04}_{-0.04}$ & $157/130$ \\
\hline
\end{tabular}
}
\tablefoot{
\tablefoottext{a} {Unabsorbed (corrected from the absorption by the ISM) average flux between 0.2 and $10\,$keV computed using the \texttt{cflux} command of \texttt{XSPEC}.}
\tablefoottext{b} {Reduced $\chi^2$ for 3 degrees of freedom.}
}
\end{table}

\section{Discussion}
\subsection{Variation of the X-ray emission}
To study the short-term variation of the X-ray emission from \stars{}, we used the Bayesian blocks method developed by \citet{scargle98} and refined by \citet{scargle13}, which creates an optimal segmentation of the source event list with blocks of constant count rate.
The event list preparation for the application of the Bayesian-block algorithm is the same as explained in \citet{mossoux14}.
The Bayesian block algorithm has to be calibrated as a function of the number of events in each observation and the desired false positive rate (i.e., the probability that a change of count rate detected by the algorithm is a false positive).
Following the method proposed by \citet{scargle13} and explained in \citet{mossoux17b}, we calibrated the algorithm by simulating 100 Poisson fluxes reproducing the observed mean count rate during the observed exposure time.
The Bayesian blocks method was then applied on the individual photon arrival times of the src+bkg and bkg event lists of each of the EPIC cameras.
The bkg contribution was then corrected by applying a weight on the Voronoi cells of the src+bkg event list as suggested by \citet{scargle13b}. No short-term variation of the X-ray emission was detected during either of the observations.
Furthermore, the total net EPIC count rates of both observations are the same within $2\sigma$.

\stars{} was previously detected during the {\it ROSAT} All-Sky Survey with a PSPC count rate of ($0.032 \pm 0.010$)\,cts\,s$^{-1}$ and an absorbed flux of $4.9 \times 10^{-13}$\,erg\,cm$^{-2}$\,s$^{-1}$ in the $0.05-2.4\,$keV energy band \citep{RASS}. 
The binary was observed from 1990 September 25 to October 10. 
The phase corresponding to the middle of this observation is $0.24 \pm 0.03$. 
Using the response matrix for the PSPC observations made before 1991 October 14 and the ancillary file computed for an extraction region of $1\arcmin$ radius centered on \stars{}, we simulated the X-ray spectrum of the best-fitting {\it XMM-Newton} spectral model as it would be observed by {\it ROSAT}. 
The resulting count rate is 0.037\,cts\,s$^{-1}$, which corresponds within the errors to the observed value quoted by \citet{RASS}.
This means that the global level of the X-ray emission from \stars{} has not changed significantly compared to its value in 1990.

\subsection{X-ray flux from the wind interaction zone}
We computed, from the revised orbital solution of \stars{}, that for an eccentricity of $e = 0.11$, the largest expected $1/d$ variation of the X-ray flux from a putative adiabatic wind interaction zone \citep{stevens92} should be around 22\% (between apastron and periastron).
These variations should only affect the part of the observed X-rays coming from the wind-wind collision and should not impact the intrinsic emission from the winds of the individual stars. 

According to the orbital phases of our {\it XMM-Newton} observations (0.40 and 0.75), the variations of the X-ray flux from the wind-wind collision between these two observations is expected to be around 7\%. 
This variation would further be diluted by the intrinsic emission from the individual stars.
Given the typical errors on the fluxes determined from the X-ray spectra (at least a few percent), it is difficult to detect such a small variation. 
The best-fit models of our spectra actually indicate a flux variation in the 0.2 -- 10\,keV range consistent with zero.
The best-fit spectral parameters change by less than 20\% between the two observations.

Using the compilation of photometric data of \citet{Reed}, we obtained $V = 6.21 \pm 0.01$ and $B-V = 0.03 \pm 0.01$ for \stars{}. 
Adopting the intrinsic colors and bolometric corrections of O6.5\,V stars from \citet{MP}, we inferred a bolometric flux of $\log{f_{\rm bol}} = -5.32 \pm 0.01$. 
Comparing this quantity with the ISM absorption-corrected X-ray flux in the 0.5--10.0\,keV band, we obtained $\log{\frac{f_{\rm X}}{f_{\rm bol}}} = \log{\frac{L_{\rm X}}{L_{\rm bol}}} = -6.76 \pm 0.04$.
This quantity is fully consistent with the canonical $\frac{L_{\rm X}}{L_{\rm bol}}$ ratio for O-type stars \citep{YN9}, and does not reveal any strong X-ray overluminosity due to the colliding wind interaction. We thus conclude that any possible excess emission from the wind interaction must be modest.

To check the validity of the hypothesis that the wind interaction in \stars{} should be in the adiabatic regime, we estimated the cooling parameter \citep{stevens92}
\begin{equation}
\chi=\frac{D}{10^7\,\mathrm{km}}\,\left(\frac{v_\mathrm{\infty,1}}{10^3\,\mathrm{km\,s^{-1}}}\right)^4\,\frac{10^{-7}\,\msun{}\, \mathrm{yr^{-1}}}{\dot{M}_1}
.\end{equation}

Using the mass-loss prescription for O-type stars of \citet{kritcka12} and assuming we have to deal with typical O6.5 main-sequence stars, we estimated mass-loss rate of $1.16$ and $1.11\times10^{-7}\,\msun{}\, \mathrm{yr^{-1}}$, respectively.
We assumed a typical terminal velocity of $v_\infty=2455\,\mathrm{km\,s^{-1}}$ for the two O6.5V stars \citep{prinja90}.

The orbital solution computed above leads to $a\,\sin{i}$ of $1742\,\rsun{}$.
From these numbers, we found that the adiabatic hypothesis is well justified in this case, since even considering the smallest orbital separation corresponding to periastron passage in the case of $i=90^{\circ}$, i.e., $D=1550.4\,\rsun{}$, the cooling parameter is $\chi =3528;$ this value is much larger than unity, meaning that the radiative cooling can be neglected in the shock region \citep{stevens92}.

To further compare the observed X-ray properties with theoretical predictions, we simulated the wind interaction region of the binary system using the analytical solution of \citet{canto96} as in \citet{rauw16b}. 

\begin{figure}
\centering
\includegraphics[trim=0.cm 0.cm 0.cm 1.5cm,clip,height=7cm]{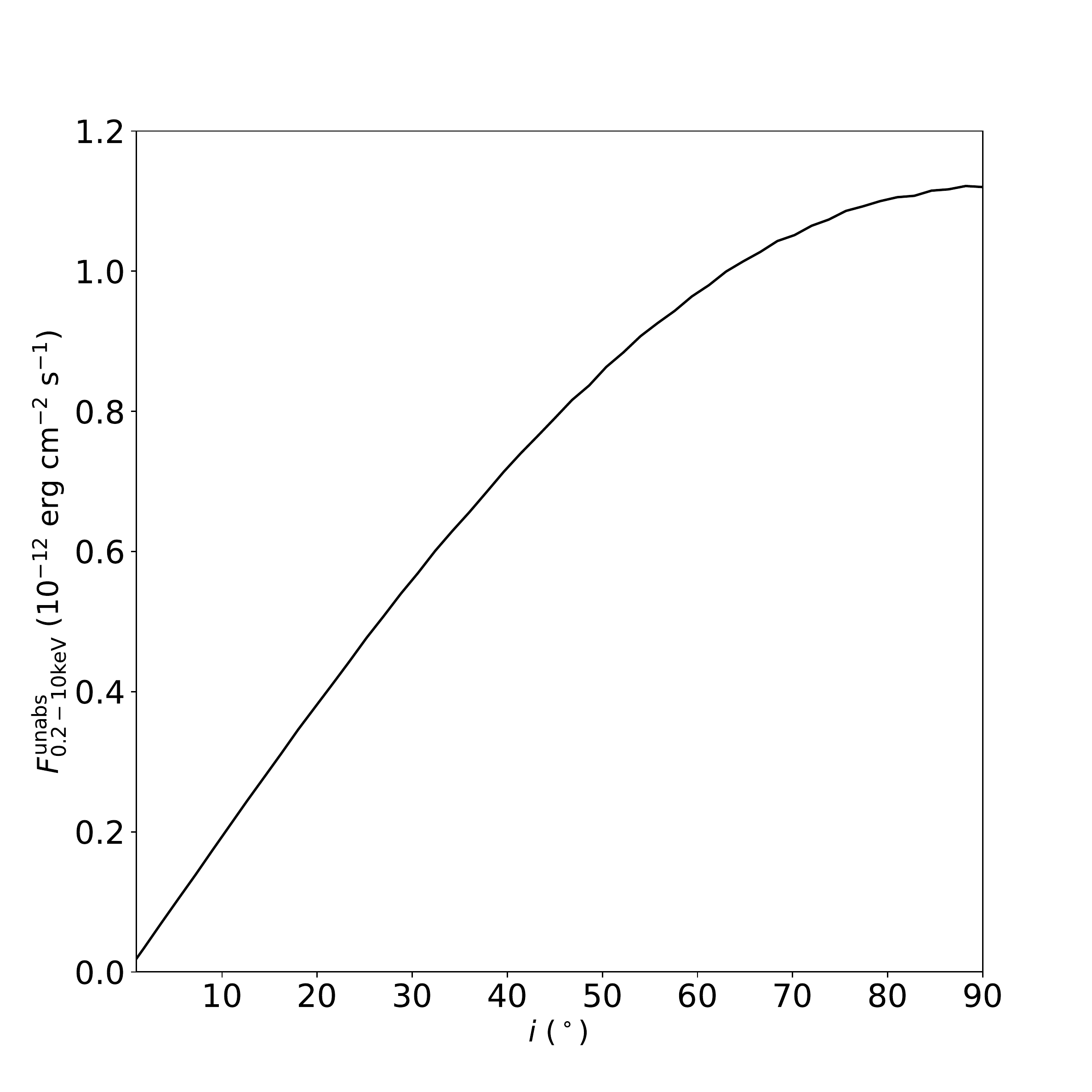}
\caption[X-ray spectra of \stars{} during the 2014 and 2016 observations]{Unabsorbed flux between 0.2 and $10\,$keV {\it vs.} the inclination of the binary.}
\label{incl}
\end{figure}

Figure~\ref{incl} shows the dependence of the theoretical unabsorbed flux with orbital inclination (solid line).
We note that the estimated unabsorbed flux is usually comparable to the total unabsorbed flux of the system quoted in Table~\ref{mcmc_table}.
We recall that the latter includes the intrinsic X-ray flux from both stars.

Moreover, the mean temperature predicted by this model is about $6.8\,$keV. This is well above the best-fitting temperature reported in Table~\ref{mcmc_table}.
We tried to fit a two-temperature model on the 2014 X-ray spectra including a component at $6.8\,$keV with a normalization to be adjusted during the fit.
The best-fitting model leads to an unabsorbed flux for the 6.8\,keV component between 0.2 and 10$\,$keV of about $1.2\times 10^{-13}\,\mathrm{erg\,cm^{-2}\,s^{-1}}$, i.e., about 11 times smaller than the flux from the intrinsic stellar emission.
Compared to Fig.~\ref{incl}, this would imply an unrealistically low orbital inclination of $i< 10^\circ$, clearly at odds with the value determined from interferometry.
The wind-shock model thus overestimates the X-ray emission for \stars{}.

This discrepancy is probably due to an overestimation of the mass-loss rate and terminal velocities of the stars.
This could be related to the ``weak-wind problem'' \citep{bouret03,marcolino09}.
\citet{kritcka12} suggested that the extreme UV (XUV) emission decreases the line driving efficiency and changes the ionization structure leading to a decay of the wind terminal velocity and mass-loss rate.
These authors computed the new relation between the mass-loss rates and luminosity but their predicted mass-loss rate still leads to too large X-ray flux.

The absence of an X-ray bright wind-wind interaction might be related to the low masses inferred above. 
However, as discussed in Sect.~\ref{rvsol}, there could be observational biases and, until future studies confirm the low masses of the stars in \stars{}, we consider this a rather speculative explanation.

\subsection{Bow shock around \stars{}}
In the literature, \stars{} is frequently considered as a runaway star. The velocity difference between a runaway star and the ISM can create a bow shock in the direction of the motion of the star.
Previous infrared  observations of \stars{} with {\it IRAS} and {\it WISE} showed the presence of an excess emission at $60\,\mu$m at about $4\arcmin$ from the binary \citep{Noriega,Peri}. 
The intensity ratio in this region is $I(60\mu\mathrm{m})/I(100\mu\mathrm{m})=0.59\pm0.04,$ which is larger than the limit of $0.3$ computed by \citet{vanburen95} to define a bow shock feature.
The presence of the bow shock thus suggests that \stars{} is a runaway binary.

\begin{figure}
\centering
\includegraphics[height=8.3cm,angle=0]{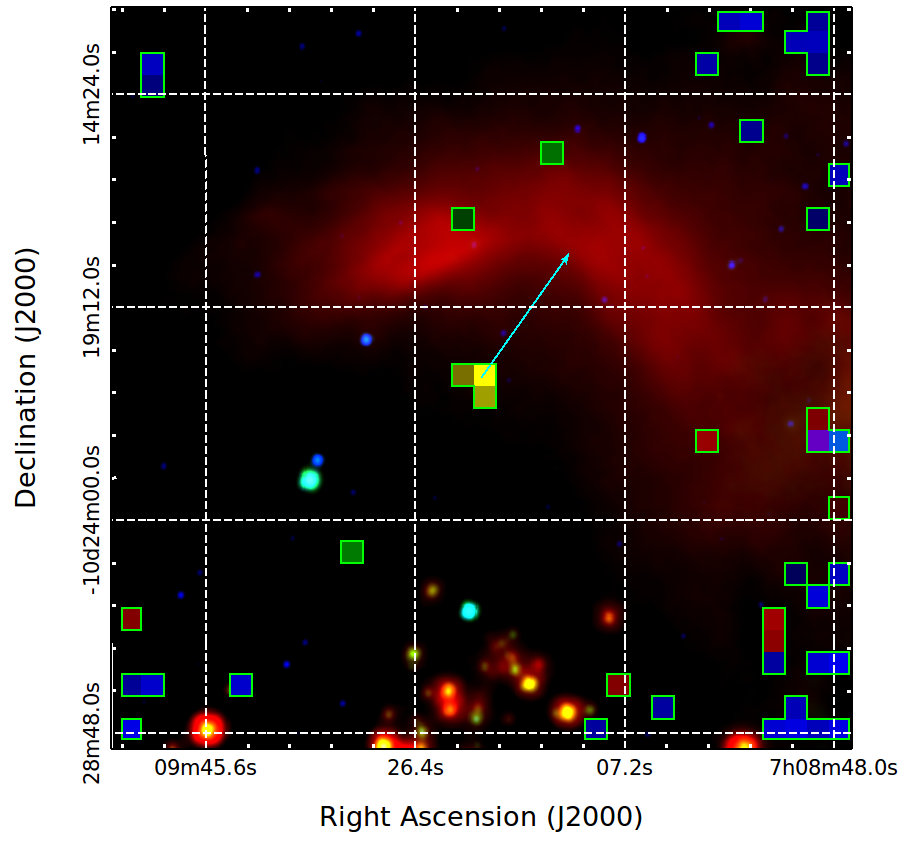}
\caption[Bow shock of \stars{}]{X-ray image (red = 180--721~eV, green = 722-6614~eV, blue = 6.6--10~keV) of \stars{} and its surroundings superimposed on the {\it WISE} image (red = $22.2\,\mu$m, green = $12.1\,\mu$m, blue = $3.4\,\mu$m).
The X-ray pixels of $30.2\arcsec \times 30.2\arcsec$ are surrounded by a green line.
The cyan arrow shows the proper motion of \stars{} taken from the Gaia Data Release 1 \citep{gaia16a}. Its size corresponds to the movement of \stars{} during $5\times 10^4$ years.}
\label{bowshock}
\end{figure}

We used the 2014 {\it XMM-Newton} observation to search for an X-ray counterpart of the bow shock.
We created the RGB image and the corresponding exposure maps using the following energy ranges: 180--721~eV (red), 722-6614~eV (green), and 6.6--10~keV (blue) leading to the same number of counts in each range.
The images were binned in order to have a minimum of one count in each pixel leading to a pixel size of $30.2\arcsec \times 30.2\arcsec$.
The RGB pixels are superimposed on the {\it WISE} image in Fig.~\ref{bowshock}.
We only display the pixels having an excess of counts larger than $2\sigma$.

The position of \stars{} is clearly observable in the soft (red and green) bands but there is no significant X-ray emission corresponding to the position of the bow shock observed in infrared. 
The only significant pixel in the bow shock region (corresponding to the location of the ``S1'' source detected by \citealt{debecker17}) is very hard, which may indicate a background source located behind the bow shock feature.

\stars{} thus joins the list of seven runaway stars or runaway candidates without detected nonthermal X-ray counterpart from the bow shock \citep{toala17}.
In this context, we also note that \citet{schultz14} failed to detect a significant $\gamma$-ray counterpart to the bow shock of \stars{}.

\begin{figure}
\centering
\includegraphics[height=8.3cm,angle=0]{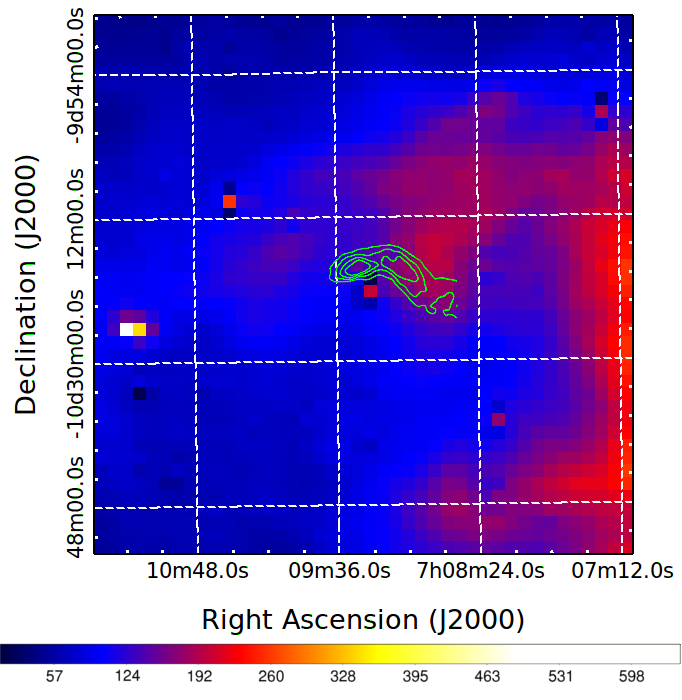}
\caption[H$\alpha$ emission from CMa OB1]{Continuum-corrected H$\alpha$ image of the CMa OB1 association observed during the Virginia Tech Spectral-Line Survey (VTSS).
The color bar is the H$\alpha$ emission in Rayleighs, i.e., in $10^6/4\pi\,\mathrm{photons/cm^2/s/sr}$.
The green lines represents the infrared emission of the bow shock as observed by WISE.}
\label{halpha}
\end{figure}

Interestingly, the position of the bow shock of \stars{} corresponds to the border of the H$\alpha$ emission from the CMa OB1 association.
Figure~\ref{halpha} shows the contour of the bow shock superimposed over the continuum-corrected H$\alpha$ image observed during the Virginia Tech Spectral-Line Survey (VTSS\footnote{\href{http://www1.phys.vt.edu/~halpha/}{http://www1.phys.vt.edu/~halpha/}}).
The bow shock thus seems to be created by the interaction between the wind of \stars{} and a local enhancement of the ionized gas of the association leading to a kind of wind-blown arc.

The systemic velocities that we inferred for the primary and secondary stars allow us to reconsider the runaway status of \stars{}. 
Adopting the formalism of \citet{Moffat} and \citet{YNPhD} along with the estimated distance of 1.2\,kpc for which we assumed a 30\% relative error, we estimated the peculiar RV of the binary system as $v_r^{\rm pec} = (10.3 \pm 3.5)$\,km\,s$^{-1}$. This is slightly lower than the value obtained by \citet{Gies}, but the latter author was not aware of the binarity of \stars{} and assumed a (constant) RV of 58.5\,km\,s$^{-1}$. 

The proper motion of \stars{}, $\mu_{\alpha} = -2.38 \pm 0.03$\,mas\,yr$^{-1}$ and $\mu_{\delta} = 3.38\pm 0.03$\,mas\,yr$^{-1}$ \citep{gaia16a}, then allows us to estimate the peculiar tangential velocity as $v_t^{\rm pec} = (40.1 \pm 11.5)$\,km\,s$^{-1}$. 
\citet{Moffat} proposed that a star should be classified as a runaway if $v_t^{\rm pec} > 42 + \sigma_{v_t^{\rm pec}}$. 
According to out findings, this is not the case of \stars{}. We also note that the peculiar RV is well below the 30\,km\,s$^{-1}$ threshold for runaways based on RVs \citep{cruz-gonzalez74}. We thus conclude that, based on the motion of the star, the runaway status of \stars{} is not established and one would rather classify it as a walk-away. 
The most compelling evidence for a relative motion compared to the ISM hence comes from the bow shock (see Fig.~\ref{bowshock}), although we have seen above that the latter might also correspond to the interface between the stellar wind of \stars{} and the ionized gas of the CMa OB1 association.

\section{Conclusions}
We studied the O-star binary system \stars{} using a set of new optical spectra and published RVs and two X-ray observations.
Applying our disentangling code on the new optical spectra, we determined RVs for the primary and secondary stars and reconstructed the spectra of the individual stars. 
Combining all existing RVs of the primary, we established an orbital period of 2103.4~days. 
Using the RVs for the primary and secondary star, we determined an orbital eccentricity of 0.11, which is surprisingly low for such a wide binary system where tidal interactions are negligible. 

Unlike our expectations, we found the X-ray emission to be extremely soft, nonvariable, and fully consistent with intrinsic emission from the components of the binary. The level of X-ray emission due to a colliding wind interaction thus appears to be much lower than expected from first principles. Either the winds interact at much lower velocities than their terminal speed or the mass-loss rates are lower than estimated from the stellar parameters.

The X-ray observations failed to reveal an X-ray counterpart for the bow shock observed in infrared. 
This bow shock is usually taken as evidence that \stars{} is a runaway star. However, we found that the peculiar radial and tangential velocities do not support this status. 

\begin{acknowledgements}
We acknowledge support through an XMM PRODEX contract (Belspo), from the Fonds de la Recherche Scientifique (FRS/FNRS), and an ARC grant for Concerted Research Actions financed by the French Community of Belgium (Wallonia-Brussels Federation). 
The TIGRE facility is funded and operated by the universities of Hamburg, Guanajuato and Li\`ege. This research has made use of the SIMBAD database, operated at CDS, Strasbourg, France. 
This work also made use of data from the European Space Agency (ESA) mission Gaia (\href{http://www.cosmos.esa.int/gaia}{http://www.cosmos.esa.int/gaia}), processed by the Gaia Data Processing and Analysis Consortium (DPAC, \href{http://www.cosmos.esa.int/web/gaia/dpac/consortium}{http://www.cosmos.esa.int/web/gaia/dpac/consortium}). 
We are grateful to Prof.\ D.J.\ Hillier for making his CMFGEN model atmosphere code available and to Dr.\ Y.\ Naz\'e for sharing the wind absorption model with us. 
We also acknowledge the Virginia Tech Spectral-Line Survey (VTSS), which is supported by the National Science Foundation, to provide the H$\alpha$ images of CMa OB1 association.
We are finally grateful to Dr~S.~Simon-Diaz for making the IACOB database available and to Dr. Virginia McSwain for a referee report that helped improve this article.
\end{acknowledgements}

\bibliographystyle{aa}
\bibliography{biblio_HD54662}

\end{document}